\documentclass[twocolumn]{aastex631}

\usepackage{mathptmx}
\usepackage[T1]{fontenc}
\usepackage{ae,aecompl}
\usepackage{graphicx}	
\usepackage{amsmath}	
\usepackage{amssymb}	
\usepackage{subfigmat}
\usepackage{enumerate}
\usepackage{multirow}
\usepackage{natbib}
\usepackage{longtable}
\usepackage{xcolor,colortbl}

\newcommand\ShortNS{\mbox{450\,$\mu$m}} 
\newcommand\LongNS{\mbox{850\,$\mu$m}}  
\newcommand\ShortS{\mbox{450\,$\mu$m} } 
\newcommand\LongS{\mbox{850\,$\mu$m} }  

\graphicspath{{./}{figures/}}

\shorttitle{JCMT Transient Survey: Six-Year Summary and Calibration 2.0}
\shortauthors{Mairs et al.}

\begin{document}

\title{The JCMT Transient Survey: Six-Year Summary of 450/850\,$\mu$m Protostellar Variability and Calibration Pipeline Version 2.0}

\correspondingauthor{Steve Mairs}
\email{drstevemairs@gmail.com}

\author[0000-0002-6956-0730]{Steve Mairs}
\affiliation{NRC Herzberg Astronomy and Astrophysics, 5071 West Saanich Rd, Victoria, BC, V9E 2E7, Canada}
\affiliation{SOFIA Science Center, Universities Space Research Association, NASA Ames Research Center, Moffett Field, California 94035, USA}
\affiliation{East Asian Observatory, 660 N. A`oh\={o}k\={u} Place,
Hilo, Hawai`i, 96720, USA}

\author[0000-0001-6324-8482]{Seonjae Lee}
\affiliation{Department of Physics and Astronomy, Seoul National University, 1 Gwanak-ro, Gwanak-gu, Seoul 08826, Korea}

\author[0000-0002-6773-459X]{Doug Johnstone}
\affiliation{NRC Herzberg Astronomy and Astrophysics, 5071 West Saanich Rd, Victoria, BC, V9E 2E7, Canada}
\affiliation{Department of Physics and Astronomy, University of Victoria, Victoria, BC, V8P 5C2, Canada}

\author[0000-0003-2353-6839]{Colton Broughton}
\affiliation{Department of Physics and Astronomy, University of Victoria, Victoria, BC, V8P 5C2, Canada}

\author[0000-0003-3119-2087]{Jeong-Eun Lee}
\affiliation{Department of Physics and Astronomy, Seoul National University, 1 Gwanak-ro, Gwanak-gu, Seoul 08826, Korea}

\author[0000-0002-7154-6065]{Gregory J. Herczeg}
\affiliation{Kavli Institute for Astronomy and Astrophysics, Peking University, Yiheyuan Lu 5, Haidian Qu, 100871 Beijing, Peoples Republic of China}
\affiliation{Department of Astronomy, Peking University, Yiheyuan 5, Haidian Qu, 100871 Beijing, China}

\author[0000-0003-0438-8228]{Graham S.\ Bell}
\affiliation{East Asian Observatory, 660 N. A`oh\={o}k\={u} Place,
Hilo, Hawai`i, 96720, USA}

\author[0000-0003-0849-0692]{Zhiwei Chen}
\affiliation{Purple Mountain Observatory, Chinese Academy of Sciences, 10 Yuanhua, 210023 Nanjing, China}

\author{Carlos Contreras-Pe\~na}
\affiliation{Department of Physics and Astronomy, Seoul National University, 1 Gwanak-ro, Gwanak-gu, Seoul 08826, Korea}

\author[0000-0001-8822-6327]{Logan Francis}
\affiliation{Department of Physics and Astronomy, University of Victoria, Victoria, BC, V8P 5C2, Canada}
\affiliation{NRC Herzberg Astronomy and Astrophysics, 5071 West Saanich Rd, Victoria, BC, V9E 2E7, Canada}

\author{Jennifer Hatchell}
\affiliation{Physics and Astronomy, University of Exeter, Stocker Road, Exeter EX4 4QL, UK}

\author{Mi-Ryang Kim}
\affiliation{Department of Physics and Astronomy, Seoul National University, 1 Gwanak-ro, Gwanak-gu, Seoul 08826, Korea}

\author{Sheng-Yuan Liu}
\affiliation{Academia Sinica Institute of Astronomy and Astrophysics, 11F of AS/NTU Astronomy-Mathematics Building, No.1, Sec. 4, Roosevelt Rd, Taipei 10617, Taiwan, R.O.C.}

\author[0000-0001-8467-3736]{Geumsook Park}
\affiliation{Research Institute of Natural Sciences, Chungnam National University, 99 Daehak-ro, Yuseong-gu, Daejeon 34134, Republic of Korea}

\author{Keping Qiu}
\affiliation{School of Astronomy and Space Science, Nanjing University Xianlin Campus, 163 Xianlin Avenue, Qixia District, Nanjing, Jiangsu, China, 210023}

\author{Yao-Te Wang}
\affiliation{Academia Sinica Institute of Astronomy and Astrophysics, 11F of AS/NTU Astronomy-Mathematics Building, No.1, Sec. 4, Roosevelt Rd, Taipei 10617, Taiwan, R.O.C.}
\affiliation{Graduate Institute of Astrophysics, National Taiwan University, No. 1, Sec. 4, Roosevelt Rd., Taipei 10617, Taiwan, R.O.C.}

\author{Xu Zhang}
\affiliation{School of Astronomy and Space Science, Nanjing University Xianlin Campus, 163 Xianlin Avenue, Qixia District, Nanjing, Jiangsu, China, 210023}

\author{The JCMT Transient Team}

\begin{abstract}
The JCMT Transient Survey has been monitoring eight Gould Belt low-mass star-forming regions since December 2015 and six somewhat more distant intermediate-mass star-forming regions since February 2020 with SCUBA-2 on the JCMT at \ShortS and \LongS and with an approximately monthly cadence. We introduce our Pipeline v2 relative calibration procedures for image alignment and flux calibration across epochs, improving on our previous Pipeline v1 by decreasing measurement uncertainties and providing additional robustness. These new techniques work at both \LongS and \ShortNS, where v1 only allowed investigation of the \LongS data. Pipeline v2 achieves better than $0.5^{\prime\prime}$ relative image alignment, less than a tenth of the submillimeter beam widths. The v2 relative flux calibration is found to be 1\% at \LongS and $<5$\% at \ShortNS. The improvement in the calibration is demonstrated by comparing the two pipelines over the first four years of the survey and recovering additional robust variables with v2. Using the full six years of the Gould Belt survey the number of robust variables increases by 50\,\%, and at \ShortS we identify four robust variables, all of which are also robust at \LongNS. The multi-wavelength light curves for these sources are investigated and found to be consistent with the variability being due to dust heating within the envelope in response to accretion luminosity changes from the central source.
\end{abstract}


\section{Introduction} 
\label{sec:intro}

The advent of sensitive large field-of-view detectors has 
launched an era of time domain astronomy with (sub)millimetre
single-dish telescopes. These data sets have been used to 
search for and characterize transient events, such as flares 
from stars \citep{mairs2019, naess21, guns21, Johnstone22} 
and relativistic jets \citep{Fuhrmann14, Tetarenko17, 
subroweit17}, as well as to monitor variability within nearby
Galactic star-forming regions associated with the stellar 
mass assembly process \citep{johnstone2018,Park19,yhlee2021}.
Over the next decade, (sub)millimetre time domain astronomy 
is anticipated to play an ever more important role, as an 
essential science mode for FYST \citep{CCAT2021}, AtLAST \citep{Ramasawmy22}, 
CMB-S4 \citep{CMB2019}, and at 
slightly shorter wavelengths for potential space-based 
far-infrared missions \citep{Andre19,Fischer19}.

Despite the significant advances in monitoring capabilities, 
calibration of (sub)millimetre observations, especially from 
the ground, remains challenging. \citet{mairs2021} analysed 
over a decade of James Clerk Maxwell (JCMT) Submillimetre 
Common  User Bolometer Array 2 (SCUBA-2) continuum imager 
observations and concluded that the peak flux uncertainty at 
\LongS after observatory-based calibrations  is 7\%, while at
\ShortS the uncertainty rises to 17\%. Notwithstanding the 
dynamic range and sensitivity of the modern large format 
(sub)millimetre detectors, these calibration uncertainties 
dominate observations of all but the faintest targets. 

To overcome this complication, the JCMT Transient Survey 
\citep{herczeg2017} developed a {\it relative} calibration 
scheme \citep{mairs2017,mairs2017GBSTrans} for \LongS SCUBA-2
observations that significantly improves, by a factor of 
three, the default calibration provided by the observatory.  
This enhanced automated reduction and relative calibration 
pipeline has allowed the team to identify many years-long 
secular protostellar variables within the Gould Belt 
\citep[upwards of 30\% of the monitored sample,][]{johnstone2018, yhlee2021}, to examine closely the
protostellar variable EC\,53 in Serpens (also known as V371\,Ser) undergoing 
episodic accretion events with an 18-month period \citep[][]{yoo2017,yhlee2020,Francis22}, and to 
investigate the months-long accretion burst associated with 
the deeply embedded protostar HOPS\,373 in Orion-B/NGC\,2068 
\citep{yoon2022}. Combined, these variability results 
obtained in the submillimetre and for optically enshrouded 
protostars have enhanced our understanding of the stellar 
mass assembly process \citep{fischer2022}.

The JCMT Transient Survey began as a three year Large Program
monitoring eight nearby Gould Belt star-forming region from 
December 2015 and has been continually extended without 
interruption, through at least  January 2024. In February 2020, six 
slightly more distant intermediate mass star-forming regions 
were added to the monitoring observations. Given the large 
accumulation of time domain data by this program,  we revisit the alignment and calibration strategy for 
the JCMT Transient Survey to ensure the best quality 
measurements. The new calibration procedure 
introduced here also allows for relative calibration of the 
SCUBA-2 \ShortS data sets. Section \ref{sec:obs} introduces 
the observations and the standard data reduction used to make
the observatory-calibrated maps at each epoch. Section 
\ref{sec:resultalign} presents the new image alignment 
procedure, while Section \ref{sec:methodlocalfluxcal} details
the updated relative flux calibration and includes an independent consistency check on its precision. Section \ref{sec:coadds} presents the
co-added images of the eight 
Gould Belt regions at \LongS and \ShortNS. We make these deep images publicly available to the astronomical community as part of this paper. We then 
present a reanalysis of source variability within the Gould 
Belt regions using the updated data sets in Section 
\ref{sec:results} and summarize the paper in Section 
\ref{sec:conclude}. 

\section{Observations and Data Reduction} 
\label{sec:obs}

The JCMT Transient Survey \citep{herczeg2017} is a James 
Clerk Maxwell Telescope Large Program (project codes: 
M16AL001 \& M20AL007) dedicated to monitoring the evolution 
of mass accretion in galactic star-forming regions. The 
survey employs the Submillimetre Common User Bolometer Array 
2 (SCUBA-2, \citealt{holland2013}), a workhorse continuum 
instrument operating simultaneously at 450 and \LongS with 
beam sizes of 9.6$^{\prime\prime}$ and  
14.1$^{\prime\prime}$, respectively. Since December 2015, 
circular  fields of $\sim30\arcmin$ usable-diameter have been
obtained approximately monthly targeting 8 star-forming 
regions in the Gould Belt: IC\,348, NGC\,1333, NGC\,2024, 
NGC\,2068, OMC\,2/3, Ophiuchus\,Core, Serpens\,Main, and 
Serpens\,South \citep[see ][ for details]{herczeg2017}. 
Combined, these fields contain more than 300 Class 0/I/Flat 
and more than 1400 Class II young stellar objects (YSOs)
as identified by the Spitzer Space Telescope 
\citep{megeath2012,stutz2013,dunham2015}. All data obtained 
since the beginning of the survey (22 December 2015) to 1 
March 2022 are included in this work (observations summarized in Appendix \ref{app:GB-summary}). 

In February 2020, the survey was expanded to monitor six 
additional fields toward regions of  intermediate/high mass 
star-formation: three fields in DR21 \citep[North: W75N, 
Central: DR21(OH), South: DR23; see][]{Schneider2010}, M17, 
M17\,SWex \citep{Povich2016}, and S255 \citep{Chavarria2008}. 
These fields are at distances less than 2\,kpc, at which the JCMT beam at \ShortS and \LongS are still capable of resolving sub-parsec dust condensations. Previously, these fields have been found to host variable young stars showing evidence of accretion outbursts \citep{2018Liu, 2019Park, 2021Chen, 2022Wenner}. These fields are representative of intermediate- to high-mass star formation regions, in which the earliest high-mass protostars are still in the making. The submillimeter monitoring observations for these fields should therefore shed light on the high-mass star-forming process. As they have only recently been added to the survey, these fields have significantly fewer observations than the 
original eight fields, requiring a separate series of 
publications for evaluation of variability. For completeness, 
a summary of the observations of these six new fields is included in Appendix 
\ref{app:HM-summary}. 

The exposure time of each JCMT Transient Survey region 
observation is adjusted based on the amount of atmospheric 
precipitable water vapour (PWV) observed along the line of 
sight to ensure a consistent background rms noise of 
$\sim12\mathrm{\:mJy\,beam}^{-1}$ at \LongS from epoch to 
epoch. The atmospheric absorption is much more severe in the 
\ShortS transmission band and the data are more susceptible 
to atmospheric variability than their \LongS counterparts. 
Typical rms noise measurements at \ShortS therefore vary over
an order of magnitude, between 100 and 
1000$\mathrm{\:mJy\,beam}^{-1}$, for observations that simultaneously yield an uncertainty of $\sim12\mathrm{\:mJy\,beam}^{-1}$ noise at \LongS (see Section 
\ref{subsec:450data}).

\cite{mairs2017} describes in detail the data reduction 
procedures used to construct individual JCMT Transient Survey
images. This work continues to make use of the configuration 
labelled as \textit{R3}, focusing on the recovery of compact,
peaked sources at the expense of accurate extended-emission 
recovery on scales larger than 200$^{\prime\prime}$. Briefly,
the {\sc{makemap}} procedure \citep{chapin2013}, part of the 
{\sc{starlink}} software suite's \citep{currie2014} 
{\sc{smurf}} package, is used to iteratively reduce the raw 
SCUBA-2 data and construct images. In order to well sample 
the beam, \ShortS maps are gridded with 
2$\arcsec$-pixels while \LongS maps are comprised of 
3$\arcsec$ pixels. A Gaussian smoothing is then performed on 
each image using a full-width at half-maximum (FWHM) 
equivalent to the angular size of 2 pixels in order to 
mitigate pixel-to-pixel noise, yielding more reliable 
peak flux measurements.  

\cite{mairs2017} also develops the first version of our 
procedures for the relative image alignment and flux 
calibration required to bring individual epochs of the same 
region into agreement, denoted \textit{Pipeline v1}, or, the 
\textit{Point-Source method}. In this work 
(\textit{Pipeline v2}), we revisit these tasks in order to 
produce more accurate and robust relative astrometry and 
calibration. Table \ref{tab:GB-obs-summary} in Appendix 
\ref{app:GB-summary} summarizes the dates, scan numbers, 
and RMS map-noise at both wavelengths, along with the 
alignment and calibration parameters for all Gould Belt 
observations. As well, to enhance the value of these JCMT 
observations, deep co-adds of the eight Gould Belt regions 
are released along with this paper (Section 
\ref{sec:coadds}).

\begin{figure*}
\plotone{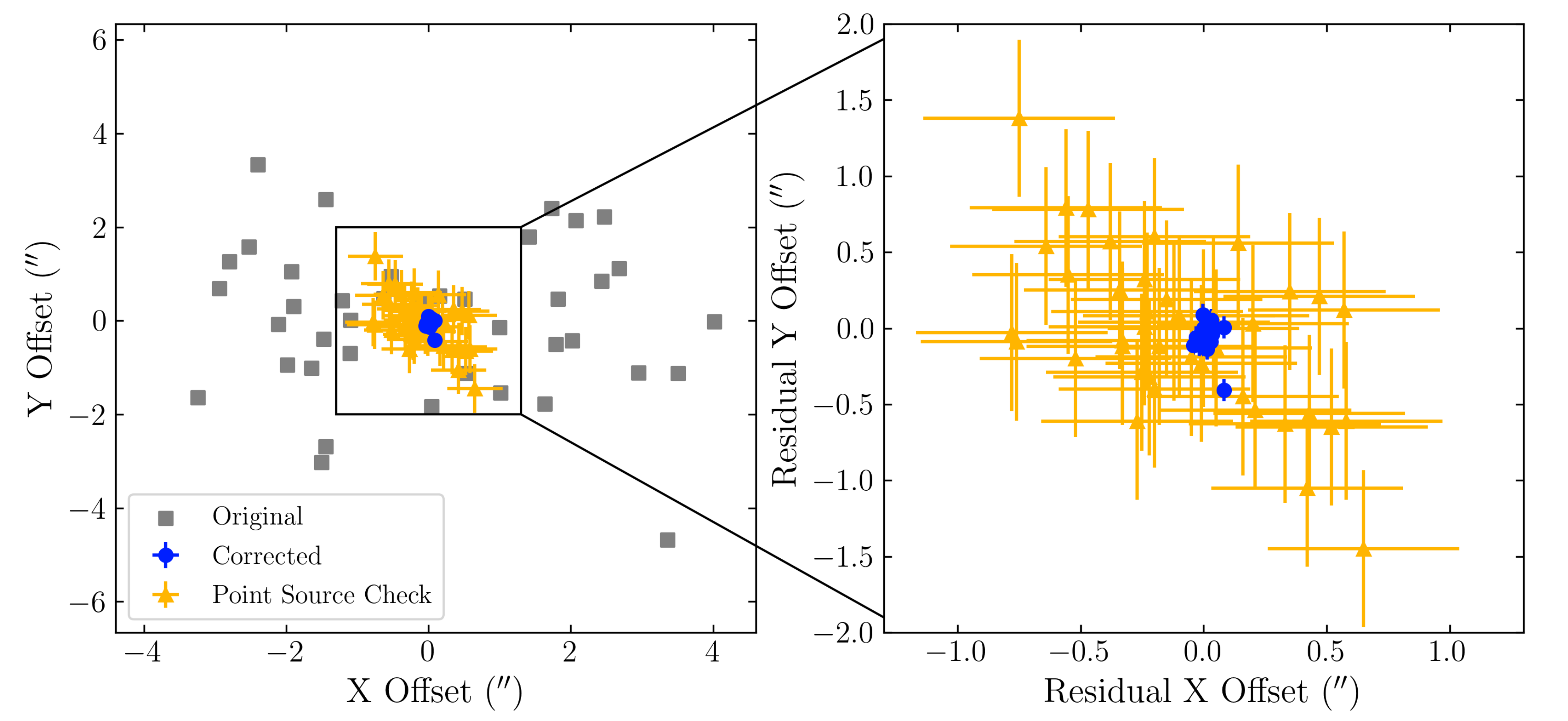}
\caption{Region: OMC 2/3. Inherent positional offsets as measured by the cross-correlation algorithm (this work) are shown as grey squares. Residual offsets after applying pointing corrections to each image are represented by blue circles. Yellow triangles show the results of a consistency check performed on the corrected images using the original (Pipeline v1; ``Point Source'') pointing-correction algorithm.}
\label{fig:Alignment}
\end{figure*}
\begin{deluxetable}{ccc}
\tablecaption{Inherent and cross-correlation algorithm-corrected uncertainty in the absolute source positions by region.}
\label{tab:Alignment}
\tablecolumns{3}
\tablewidth{0pt}
\tablehead{
\colhead{Region} &
\colhead{($\Delta$R.A.,$\Delta$Dec)$_{\mathrm{inherent}}$ ($\arcsec$)}  &
\colhead{($\Delta$R.A.,$\Delta$Dec)$_{\mathrm{corrected}}$ ($\arcsec$)}
}
\startdata
IC 348        & (1.2, 2.2) & (0.3, 0.2) \\
NGC 1333      & (1.5, 2.2) & (0.3, 0.5) \\
NGC 2024      & (1.5, 1.8) & (0.8, 0.8) \\
NGC 2068      & (1.5, 1.9) & (0.3, 0.4) \\
OMC 2/3       & (2.0, 1.6) & (0.4, 0.5) \\
Oph Core      & (2.2, 1.2) & (0.4, 0.6) \\
Serpens Main  & (1.4, 2.0) & (0.5, 0.5) \\
Serpens South & (1.2, 1.6) & (0.3, 0.3) \\
\enddata
\end{deluxetable}

\section{Image Alignment}
\label{sec:resultalign}

The JCMT has an inherent pointing uncertainty of 
$2^{\prime\prime}-4^{\prime\prime}$ (see \citealt{mairs2017} 
and Figure \ref{fig:Alignment}). This pointing offset is the 
same at both 450 and \LongS because both focal planes have 
the same field of view on the sky and observations are 
carried out simultaneously by means of a dichroic 
beam-splitter \citep{chapin2013}.  Relative image alignment can therefore be achieved by determining positional offsets between epochs using only the \LongS images, for which the background measurement noise is consistent over time (see 
Section \ref{sec:obs}).

As part of the JCMT Transient Survey data reduction process for relative alignment, the original ``Point 
Source'' method \citep{mairs2017} tracked a set of bright, 
compact, peaked sources identified in each epoch and compared
their measured peak positions to the first image obtained for
that region. For each additional epoch, the calculated 
average offsets in both R.A. and Declination over this set of
sources were used to correct the newly obtained data to a 
fixed grid \citep{mairs2017}.  While this method produced 
maps in relative alignment to within a factor of 
$\sim1^{\prime\prime}$, the technique requires at least 
several bright, point-like sources to be present throughout 
the map and the derived offsets are subject to Gaussian 
fitting uncertainties introduced by pixel-to-pixel noise. 
Furthermore, this method did not attempt to remove the 
pointing uncertainty of the first epoch.

A more robust \textit{Pipeline v2} 
algorithm has been developed and is now included in the updated data
reduction pipeline. This new technique cross-correlates the 
reconstructed images associated with each epoch, prior to 
Gaussian smoothing, to determine the relative alignment 
offsets and estimates the absolute pointing from the 
statistics of the individual pointings, assuming that there 
is no systematic pointing offset at the telescope. In detail,
the new image alignment procedure defines a nominally 
$20\arcmin\times20\arcmin$ sub-field centered on the middle 
of a given map\footnote{In two of the 14 fields, DR\,21C and 
DR\,21N, the analysed region is slightly shifted from the 
center in order to include bright and structured emission 
that would otherwise be cut off.} taken at epoch $i$, 
$\mathrm{Map}_{i}$,  and cross-correlates the information 
contained within against the same sub-field extracted from 
the first observation of the given region, 
$\mathrm{Map}_{0}$. The method employs {\sc scipy}'s 
 2-D, discrete Fourier transform algorithm, 
{\sc fft2} \citep{SciPy}:
\begin{equation}
   XC = \mathcal{F}^{-1}[\mathcal{F}(\mathrm{Map}_{i})\mathcal{F}(\mathrm{Map}_{0})^{*}]
\end{equation}
where $XC$ is the computed cross-correlation function, 
$\mathcal{F}$ represents the 2-D discrete Fourier transform, 
and $^{*}$ represents the 
complex conjugate. $\mathrm{Map}_{i}$ and $\mathrm{Map}_{0}$ 
contain the same structured emission at slightly different 
positions, tracing the 
inherent pointing uncertainty between the epochs. Meanwhile, 
the pixelated random measurement noise is uncorrelated 
between epochs and therefore
does not produce a localized peak in $XC$. The epoch offset 
is measured by fitting a Gaussian function to $XC$ with a 
Levenberg-Marquardt Least 
Squares algorithm. The necessary offsets that must be applied
to $\mathrm{Map}_{i}$ in order to bring it into relative 
alignment with 
$\mathrm{Map}_{0}$ are derived by measuring the misalignment 
of the peak of the cross-correlation function from the center
in both the R.A.\ and
Declination directions. In the final step, the central 
coordinates of $\mathrm{Map}_{i}$ are also uniformly shifted to 
the median R.A.\ and 
Declination measured over the \textit{unaligned maps} 
observed on or before 15 June, 2021 as a reasonable 
estimate of the true 
astrometry assuming pointing errors over many epochs are unbiased. 
The original \textit{Point-Source method} continues to be used as a 
consistency check for this more robust \textit{Pipeline v2} 
algorithm.

The left panel of Figure \ref{fig:Alignment} shows the 
original telescope pointing offsets with respect to the 
derived true astrometry toward 
Orion OMC\,2/3, as measured by the $XC$ method, for all 
epochs (grey squares), the residual offsets measured after the 
cross-correlation alignment 
corrections (\textit{Pipeline v2}) have been applied (blue 
circles), and the residual \textit{Point-Source method} 
alignment consistency checks 
after the cross-correlation alignment corrections have been 
applied (yellow triangles). The right panel of Figure 
\ref{fig:Alignment} shows a 
zoomed-in comparison between the residual offsets measured by
each method. There is excellent agreement between the new and
old techniques. The 
standard deviation of the residual cross-correlation offsets 
suggests the maps are self-consistently aligned to better 
than $0.2^{\prime\prime}$. The point-source verification 
suggests the alignment uncertainty is better than 
$\sim0.5^{\prime \prime}$, though the inherent uncertainty using the point-source
method is larger than for the cross-correlation method. Table
\ref{tab:Alignment} 
shows the alignment results for each of the eight Transient 
Survey Gould Belt regions. The $\Delta$R.A. and $\Delta$Declination
values are derived by 
calculating the standard deviation of the measured map 
offsets from the median (defined to be 0,0). The 
``corrected'' residual offsets in the 
table refer to the point-source alignment verification 
(yellow triangles in Figure \ref{fig:Alignment}) which 
provide a more conservative  
measurement of the corrected pointing uncertainty than the 
cross-correlation self-consistency check. Derived pointing 
offsets for each Gould
Belt observation are given in Table \ref{tab:GB-obs-summary} 
in Appendix \ref{app:GB-summary}.

\begin{figure*}[t]
\plottwo{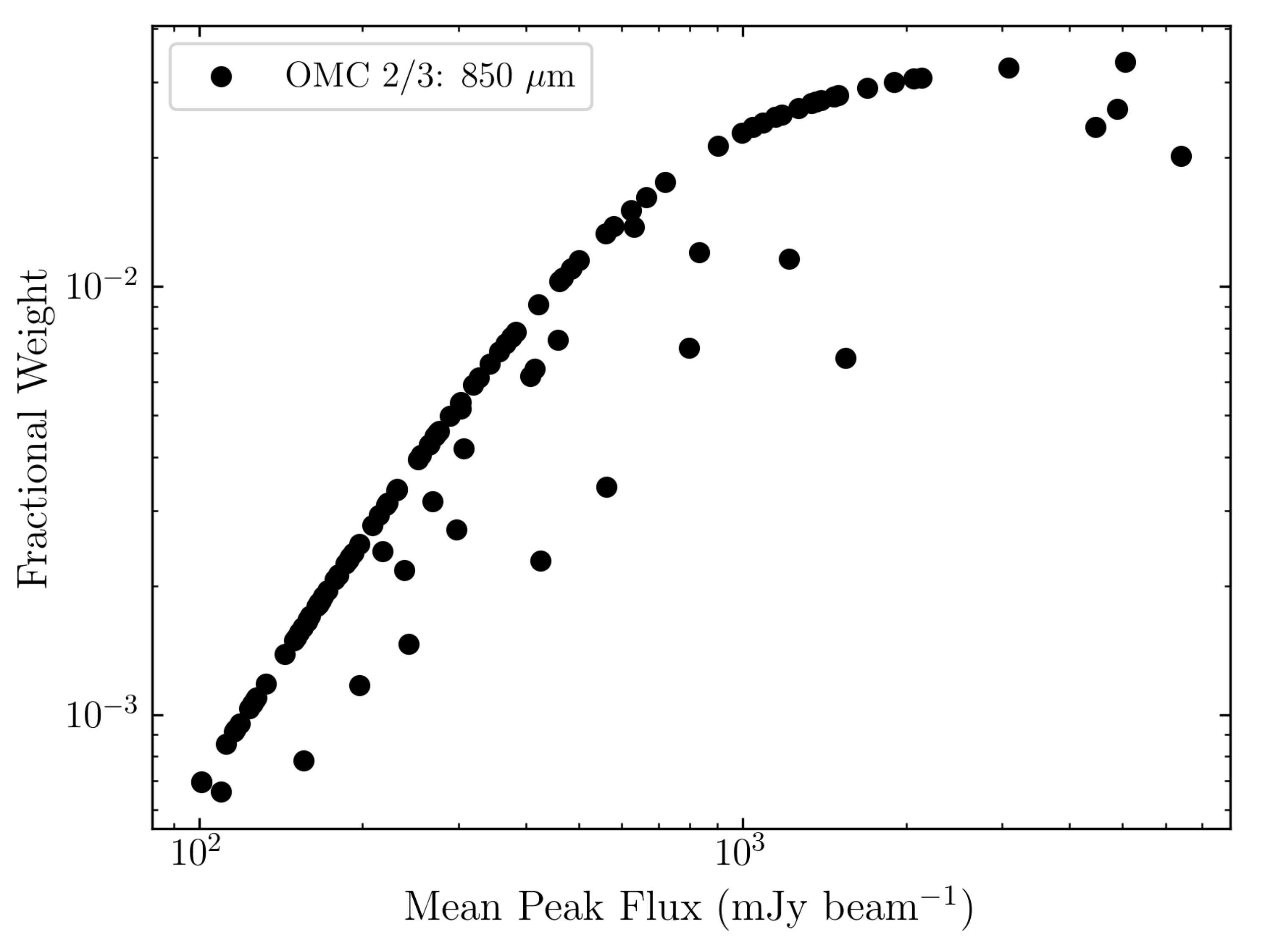}{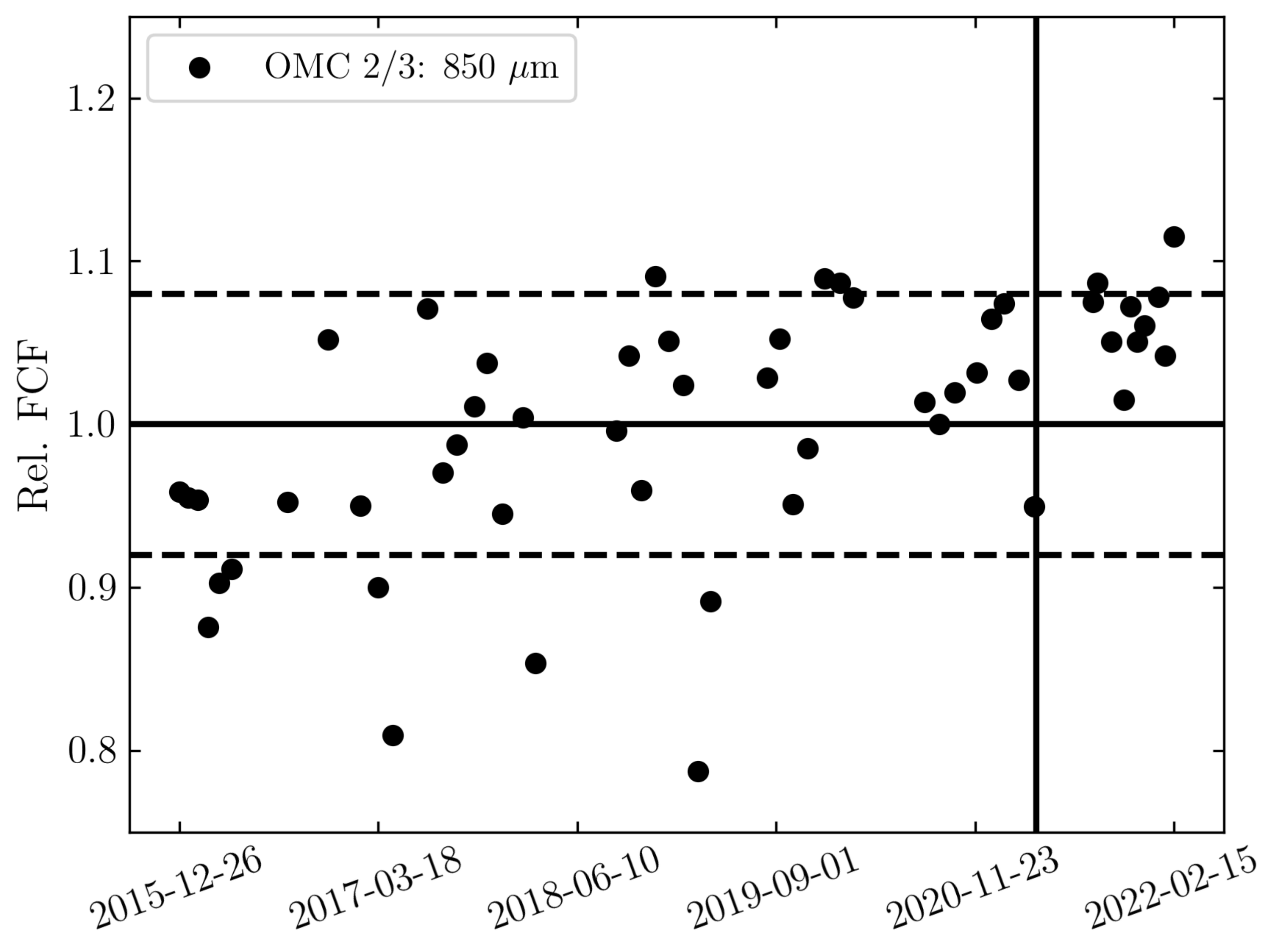}
\caption{Region: OMC 2/3. Left: The fractional weight each identified point-like source carries in the weighted, relative flux calibration. Right: Calculated Relative FCFs as a function of time. The vertical line denotes the date before which the median flux is calculated for each source to employ in the relative normalisation (UTC 2021-04-10; see point 2 in Section \ref{sec:weightedsourcecal}). The dashed horizontal lines show the one sigma range of the Relative FCF values.}
\label{fig:WeightedCalSum}
\end{figure*}

\section{Relative Flux Calibration}
\label{sec:methodlocalfluxcal}

Both the 450 and 850\,$\mu$m images are flux-calibrated in a 
relative sense over time by measuring peak fluxes of bright, 
compact submillimetre emission sources that were originally 
catalogued in co-added images of each target field. The 
{\sc{Fellwalker}} \citep{berry2015} source identification 
algorithm was used to find the locations of compact, peaked 
sources; it is part of {\sc{Starlink}}'s \citep{currie2014} 
{\sc{Cupid}} \citep{CUPID2013} software package. Sources 
below brightness thresholds of 
$1000\mathrm{\:mJy\:beam}^{-1}$ at \ShortS and 
$100\mathrm{\:mJy\:beam}^{-1}$ at \LongS are excluded from 
this final source calibration position catalogue. Furthermore, sources that 
were previously identified as known submillimetre variables 
by \citet{mairs2017GBSTrans}, \citet{johnstone2018}, and 
\citet{yhlee2021} are removed.

The peak flux of each source in a given catalogue is measured
in each observed epoch to construct initial light curves for 
all identified objects. Prior to measuring peak flux, the 
maps used are Gaussian-smoothed by 2 pixels to better match 
the beam, and are relatively aligned to much better than the 
scale of a single pixel (see Sections \ref{sec:obs} and 
\ref{sec:resultalign}). The peak flux in each epoch is then 
accurately measured at a fixed pixel location, without 
requiring a Gaussian fit for each epoch. 

The ``Fiducial'' (expected) standard deviation, 
$\mathrm{SD_{\mathrm{fid}}}$, in the light curve of a given 
source, $i$, has the form \citep{johnstone2018}:
\begin{equation}
    \label{eq:SDfid}
    \mathrm{SD_{\mathrm{fid}(i)}} = [(\mathrm{n}_{\mathrm{RMS}})^{2} + (\sigma_{\mathrm{FCF}}\times f_{m}(i))^{2}]^{1/2}, 
\end{equation}
where 
$\mathrm{n}_{\mathrm{RMS}}$
is the typical RMS noise measured across the epochs (dominating 
faint sources), $\sigma_{\mathrm{FCF}}$
is the expected relative flux-calibration uncertainty
that can be achieved by the algorithms described below 
(dominating bright sources), and $f_{m}(i)$ is the mean 
peak flux of source $i$. At \LongS and using the original \textit{Pipeline v1} calibration approach, \citet{johnstone2018} 
found $\mathrm{n}_{\mathrm{RMS}} = 12\mathrm{\:mJy\:beam}^{-1}$ and $\sigma_{\mathrm{FCF}} = 0.02$ (that 
is 2\%). 

\subsection{Iterated Weighted Source Calibration}
\label{sec:weightedsourcecal}

The \textit{Pipeline v2} relative flux calibration proceeds as follows. 
Once the initial light curves for all sources in a given 
region are constructed, an iterative algorithm is used to 
weight each source either as a favorable or unfavorable 
\textit{calibrator source} based on the robustness of each 
observed map and the constancy of the source light-curve over
time. With this information in hand, the full set of 
calibration sources are used to derive a ``Relative Flux 
Calibration Factor'' (Relative-FCF; $R_{\mathrm{FCF}}$) for each epoch. 
The $R_{\mathrm{FCF}}$ is a multiplicative constant with which to 
multiply a given epoch in order to bring the map into 
agreement with the weighted mean flux brightnesses over all 
sources, as measured in the co-added image of all data 
obtained on or before 10 April 2021. The details of the 
algorithm are as follows:

\begin{enumerate}
    \item Initial $R_{\mathrm{FCF}}$ and their associated uncertainties, $\sigma_{\mathrm{FCF}}$, are set to be 1 and 2\%, respectively, for each epoch and each wavelength. Using these initial estimates, an iterative process begins.

    \item For each source, excluding known variables, the weighted mean flux, $\bar{f}_\mathrm{source}$ is estimated as follows:

    \begin{equation}
        \bar{f}_\mathrm{source} = \frac{\sum_{e=1}^{N_{e}}R_{\mathrm{FCF},e} \times f_{\mathrm{source},e}/\sigma_{\mathrm{FCF},e}^{2}}{\sum_{e=1}^{N_{e}}1/\sigma_{\mathrm{FCF},e}^{2}},
    \end{equation}

    where $e$ represents each epoch, $N_{e}$ represents the number of epochs, and $f_{\mathrm{source},e}$ represents the flux of the given source in epoch $e$.

    \item The source uncertainty, $\sigma_{\mathrm{source}}$ is then estimated by calculating the standard deviation in the source flux across all epochs, weighted by the epoch uncertainty. If the calculated source standard deviation is smaller than the fiducial standard deviation (Equation \ref{eq:SDfid} for 850\,$\mu$m and 5\% of the mean flux at 450\,$\mu$m), the fiducial value is adopted in order to prevent a runaway effect wherein the brightest, most stable sources obtain outsize weights.
    
    \item Source flux thresholds of $1000\mathrm{\:mJy\:beam}^{-1}$ at \ShortS and 
$100\mathrm{\:mJy\:beam}^{-1}$ at \LongS are then applied to excise sources that are too faint to provide a meaningful contribution to the overall flux calibration.

    \item The $R_{\mathrm{FCF}}$ and formal uncertainty for each individual epoch are then determined by applying a source-weighted linear least squares fit between the peak flux measurements in that observation and their weighted mean values calculated over all observations taken before UTC 2021-04-10. The intercept is fixed at the origin and the calculated slope and its standard deviation yield the best-fit $R_{\mathrm{FCF}}$ and its \textit{formal} uncertainty, $\sigma_{\mathrm{FCF,formal}}$. In the equations below, $s$ represents a given source and $e$ represents a given epoch:

    \begin{equation*}
        R_{\mathrm{FCF,e}} = \frac{\sum_{s=1}^{N_{\mathrm{sources}}}f_{\mathrm{s},e}\times\bar{f}_\mathrm{s}/\sigma_{\mathrm{s}}^{2}}{\sum_{s=1}^{N_{\mathrm{sources}}}f_{\mathrm{s},e}^{2}/\sigma_{\mathrm{s}}^{2}}
    \end{equation*}

    \begin{equation*}
        \sigma_{\mathrm{FCF,formal,e}} = 1/\sum_{s=1}^{N_{\mathrm{sources}}}f_{\mathrm{s},e}^{2}/\sigma_{\mathrm{s}}^{2}
    \end{equation*}

    \item The $R_{\mathrm{FCF}}$ uncertainties ($\sigma_{\mathrm{FCF}}$) are estimated by 
    calculating the standard deviation in source fluxes measured within the given epoch normalized to their respective mean fluxes across all epochs. Similarly to Step 3, an empirical analysis yielded a minimum threshold for $\sigma_{\mathrm{FCF}}$ of 70\% of the \textit{formal} $R_{\mathrm{FCF}}$ uncertainty to prevent a runaway events. Therefore, if $\sigma_{\mathrm{FCF}}$ becomes less than 70\% of the \textit{formal} $R_{\mathrm{FCF}}$ uncertainty, $\sigma_{\mathrm{FCF}}$ is set to $0.7\times\sigma_{\mathrm{FCF,formal}}$.

    \item Using the newly calculated Relative-FCFs, $R_{\mathrm{FCF}}$, and their associated uncertainties, $\sigma_{\mathrm{FCF}}$, repeat Steps 2 through 6 until convergence (achieved in at most 5 iterations). Over time, if a new variable source emerges, it will be automatically down-weighted in the calibration and identified through its change in fractional weight as in the left panel of Figure \ref{fig:WeightedCalSum} or through the calibration consistency check, described below in Section \ref{subsec:consistency}.
    
\end{enumerate}

If all sources have the same fractional 
uncertainty then they each contribute equal weight. For 
non-varying bright sources this is expected (see Eqn.\ \ref{eq:SDfid}) as the calibration dominates the peak flux 
uncertainty. Alternatively for faint sources, dominated by a
constant noise, the relative weighting varies directly as the
source brightness. Occasionally sources have additional 
measurement uncertainties due to location in the map,  
neighbor crowding, or potential variability that has not yet
been confirmed (and hence the source has not yet been removed
from the list). Due to the increased uncertainty, these 
sources automatically are down-weighted by the algorithm (see point 7, above). All known non-varying protostars are retained in the pool of calibrators, provided the
submillimetre flux is above the  thresholds. In this way, deviations in the calibration results over time are used to identify newly varying YSOs and previously undiscovered YSOs such as deeply embedded protostars that were historically too faint to be included
in previous surveys.

Taking the OMC\,2/3 region as an example, in the left panel 
of Figure \ref{fig:WeightedCalSum} the upper envelope follows
the expected weighting relation, with a few sources lying 
lower due to higher-than-expected uncertainties in their 
fluxes signaling potential variability or difficulty in 
measuring a reliable peak flux value.  The right panel of 
Figure \ref{fig:WeightedCalSum} plots the resultant 
$R_{\mathrm{FCF}}$ required for each individual epoch. The trend 
upward in the right panel is seen across all Gould Belt 
regions and corresponds to systematic changes in the JCMT flux
throughput since 2015 which changed the Standard-FCFs published by the observatory, described in detail by 
\cite{mairs2021}. These changes were: 1. a filter stack 
replacement in November 2016, and, 2. secondary mirror unit 
maintenance that improved the beam profiles in June 2018. The
flux conversion factors presented in \cite{mairs2021} are 
reciprocals of those derived in this work and thus decrease 
over time while the $R_{\mathrm{FCF}}$ presented here appear to 
increase. The magnitude of the Standard-FCF changes agree in both of these
studies. Derived $R_{\mathrm{FCF}}$ values for each observation are given
in Table \ref{tab:GB-obs-summary}, in Appendix 
\ref{app:GB-summary}.

\begin{figure}
\plotone{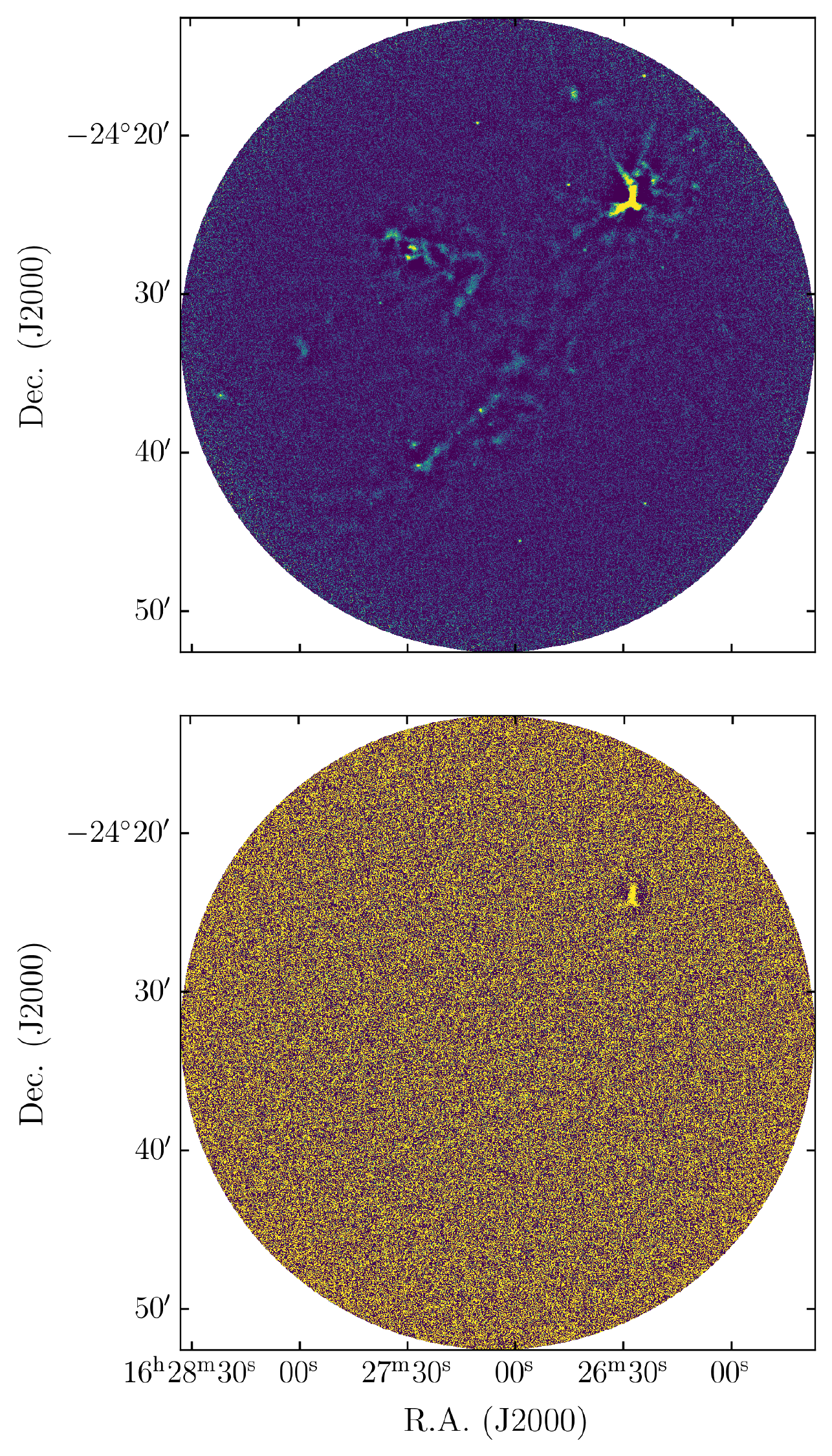}
\caption{450\,$\mu$m observations of the \mbox{Ophiuchus Core} field obtained in dry weather conditions $\tau_{225} = 0.038$,\ $\mathrm{Airmass=1.51}$ on 17 April 2016 (top) and ``wet'' weather conditions $\tau_{225} = 0.11$,\ $\mathrm{Airmass=1.64}$ on 21 May 2020 (bottom). The images share a common color map and scaling.}
\label{fig:goodbadweather}
\end{figure}
\begin{figure*}
\plottwo{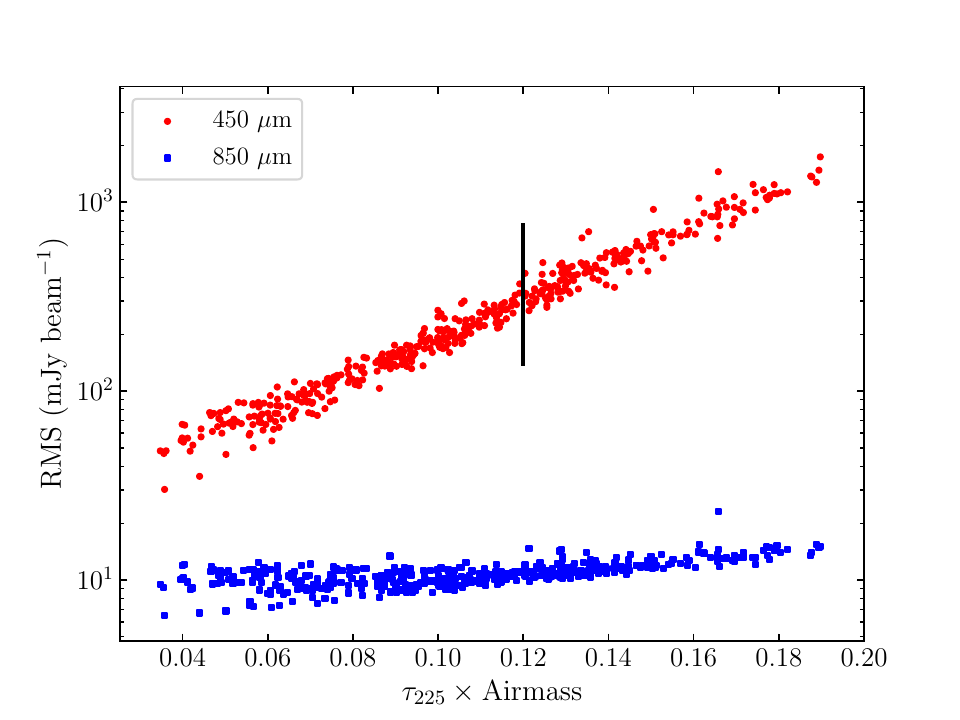}{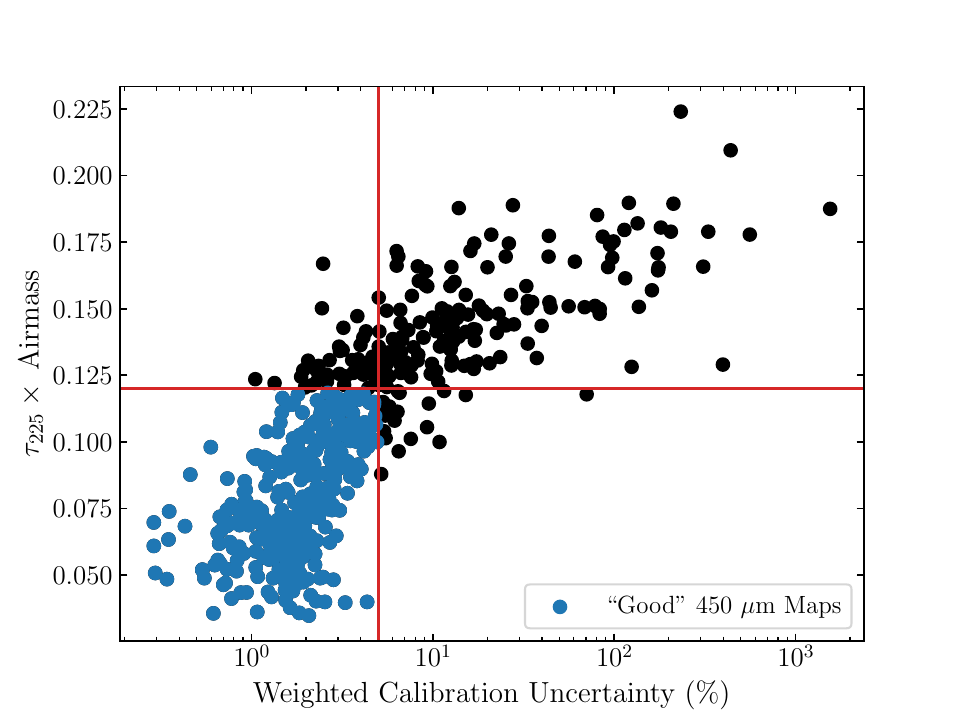}
\caption{\textit{Left:} 450 (red) and 850$\,\mu$m (blue) RMS versus $\tau_{225}\times\mathrm{Airmass}$ (a proxy for the atmospheric transmission). The vertical black line indicates the $\tau_{225}\times\mathrm{Airmass}$ threshold as indicated in the right panel. \textit{Right:} $\tau_{225}\times\mathrm{Airmass}$ versus the weighted calibration factor uncertainty (\textit{Pipeline v2}). The Blue datapoints indicate the ``usable'' maps as described in Section \ref{subsec:450data}.}
\label{fig:RMSvsTransmission}
\end{figure*}

\subsection{Selecting Reliable 450 micron data}
\label{subsec:450data}

As previously discussed, the \LongS observations obtained by 
the JCMT Transient Survey have a consistent background RMS 
noise of $\sim12\mathrm{\:mJy/beam}$. The \ShortS data, 
however, are much more sensitive to changes in atmospheric 
water vapour and airmass, causing more than an order of 
magnitude of spread in background RMS values over the 
duration of the survey (see Figures \ref{fig:goodbadweather} 
and \ref{fig:RMSvsTransmission}). Therefore, in conditions of
poor atmospheric transmission, even the brightest sources in 
\ShortS maps will be overwhelmed by the noise and no reliable
peak flux measurements can be obtained. In order to identify
the usable \ShortS maps, in the right panel of Figure 
\ref{fig:RMSvsTransmission} we plot a proxy for the 
atmospheric transmission, the opacity of the atmosphere 
measured at \mbox{225 GHz} ($\tau_{225}$) multiplied by the 
airmass of the observation, against the calculated 
Relative-FCF uncertainty ($\sigma_{\mathrm{FCF}}$ ) for each epoch and identify a box 
within which the brightest \ShortS peak flux values have a 
sufficient signal-to-noise ratio to return an accurate flux 
calibration factor. The \ShortS map is defined as ``usable'' 
if:

\begin{enumerate}
    \item $\tau_{225}\times\mathrm{Airmass}<0.12$
    \item $\mathrm{Relative\:Flux\:Calibration\:Uncertainty} < 5\%$
\end{enumerate}
Table \ref{tab:GB-obs-summary}, in Appendix 
\ref{app:GB-summary} includes the derived \ShortS $R_{\mathrm{FCF}}$ values for those epochs which satisfy the above conditions.  The table also allows for a calculation of the typical \ShortS RMS noise in an epoch, $\mathrm{n}_{\mathrm{RMS}} = 130 \pm 65 \mathrm{\:mJy\:beam}^{-1}$, which can be used in Equation \ref{eq:SDfid} to estimate the ``Fiducial" \ShortS standard deviation of any particular monitored source.

\subsection{Relative Flux Calibration Consistency Check}
\label{subsec:consistency}

An independent relative flux calibration 
algorithm is employed to verify the consistency of the 
weighted peak flux calibration scheme described above. The 
method is based on the original \textit{Point-Source method} 
described by \cite{mairs2017}, which requires identifying 
\textit{families} of bright, reliable (non-varying) compact 
sources and constructing light curves (flux versus time). The
\textit{normalized} light curves of the non-varying 
\textit{family} sources trace the inherent JCMT calibration 
uncertainty from epoch to epoch, i.e. if a given map is 
brighter than the mean, the average deviation from the mean 
of the family-source peak fluxes will yield a flux correction
factor for that epoch. The standard deviation in 
normalised peak fluxes representing the uncertainty.
These results are not used to perform calibration of the data sets. This method is used only as a confirmation of the precision of the
weighted peak flux calibration scheme.

This verification step does not select reliable \ShortS maps in the same manner as described above. Instead, \ShortS maps are individually selected based on the theoretical $R_{\mathrm{FCF}}$ uncertainty that might be achieved given the brightness of the sources present in the map that could constitute a \textit{family} along with the RMS background noise of the map itself. At \ShortNS, the target $R_{\mathrm{FCF}}$  uncertainty is defined to be 5\% to match the weighted flux calibration threshold. The theoretical relative flux 
calibration uncertainty 
($\sigma_{\mathrm{FCF,theory}}$) produced by a given \textit{family} of calibrator sources is calculated by
\begin{equation}
\sigma_{\mathrm{FCF,theory}} = \frac{1}{\mathrm{SNR}_{\mathrm{ensemble}}},
\end{equation}
where,
\begin{equation}
\mathrm{SNR}_{\mathrm{ensemble}} = \frac{f\times\sqrt{N}}{\mathrm{RMS}}.
\end{equation}
Here $f$ is the peak-flux of the faintest source considered in $\mathrm{mJy/beam}$, $N$
is the number of sources with peak-fluxes greater than or equal to $f$, and $\mathrm{RMS}$ is the root mean square background noise of a given map in $\mathrm{mJy/beam}$. Therefore, in order to achieve a relative flux calibration 
uncertainty better than $\sim$5\%, the RMS background noise threshold for 
whether to consider an individual epoch as reliable is:
\begin{equation}
\label{eq:RMSThresh}
\mathrm{RMS} \le 5\%\times f\times\sqrt{N}\mathrm{\:\:mJy/beam}.
\end{equation}

To identify the trustworthy \textit{families} of sources for each region, 
the same brightness thresholds for source consideration as in the 
weighted flux-calibration scheme described above, $1000\mathrm{\:mJy\:beam}^{-1}$ at \ShortS and 
$100\mathrm{\:mJy\:beam}^{-1}$ at \LongNS, are initially employed. Unlike in the weighted calibration method, however, not all of these sources will be used to verify the map calibration, they are simply used as the pool of \textit{potential} calibrator 
sources. These \textit{potential} calibrators are arranged in ascending flux order and known variables identified by 
\cite{mairs2017GBSTrans,johnstone2018}, and 
\cite{yhlee2021} are removed. Beginning with the brightest 
source and sequentially including fainter sources one-by-one 
(excluding known variables), ever larger potential \textit{families} are defined. For each potential \textit{family}, Equation \ref{eq:RMSThresh} is used to determine which maps achieve the RMS threshold to produce a theoretical $R_{\mathrm{FCF}}$ uncertainty of 5\%. At \LongNS, all epochs are included in the analysis. At \ShortNS, maps obtained in poor atmospheric transmission conditions are excluded. The set of reliable maps in each region generally match those identified using the method described in Section \ref{subsec:450data} with a few exceptions (see Appendix \ref{app:GB-summary}).

At both wavelengths, the potential calibrator groups are then
narrowed down by optimising three key parameters: (1.) the 
brightest set of sources that (2.) return the lowest 
$\sigma_{\mathrm{FCF}}$ using (3.) the largest number 
of maps. The final point applies specifically to \ShortS data. 
For more detail regarding the trade-off between a higher number
of sources and a lower $\sigma_{\mathrm{FCF}}$, see
Section 4.2 of \cite{mairs2017}.

To test the $\sigma_{\mathrm{FCF,theory}}$ values, 
light curves of each potential calibrator group member (in 
the narrowed-down subset) were constructed using only the 
maps that met the RMS threshold criteria described above for 
the faintest source in the group. As for the iterative, 
weighted calibration method, the light curve fluxes are 
measured at the known peak pixel location of each source at 
each aligned epoch. The light curves are then normalised by 
dividing each flux measurement by the median for that source 
across the fixed set of observations. 

To select the final calibrator \textit{family} (the group of 
sources that will be used to relatively flux calibrate the 
data; see \citealt{mairs2017}), the sources in the 
narrowed-down potential calibrator group are paired with one 
another in every possible combination for a given family size
from 2 sources to the number of sources in the group. For 
each pair of sources in each family size, the normalised 
fluxes are compared to one another, epoch by epoch. The 
standard deviation of this normalised flux ratio is an 
indication of how well the sources agree with one another as 
they track the inherent JCMT flux calibration uncertainty 
from epoch to epoch. Higher standard deviation values 
indicate the sources are in less agreement. The calibrator 
family is selected by optimizing the group size with the 
minimum standard deviation threshold of each group. Finally, 
in each epoch, this secondary $R_{\mathrm{FCF}}$ is used to verify the main weighted calibration method, taking the 
average normalised flux value among the sources comprising 
the optimized family. The normalization is calculated over 
the same time period as the \textit{Pipeline v2}, data taken 
prior to 2021-04-10 UTC such that the solutions are fixed for
future observations.

\begin{figure*}
\plotone{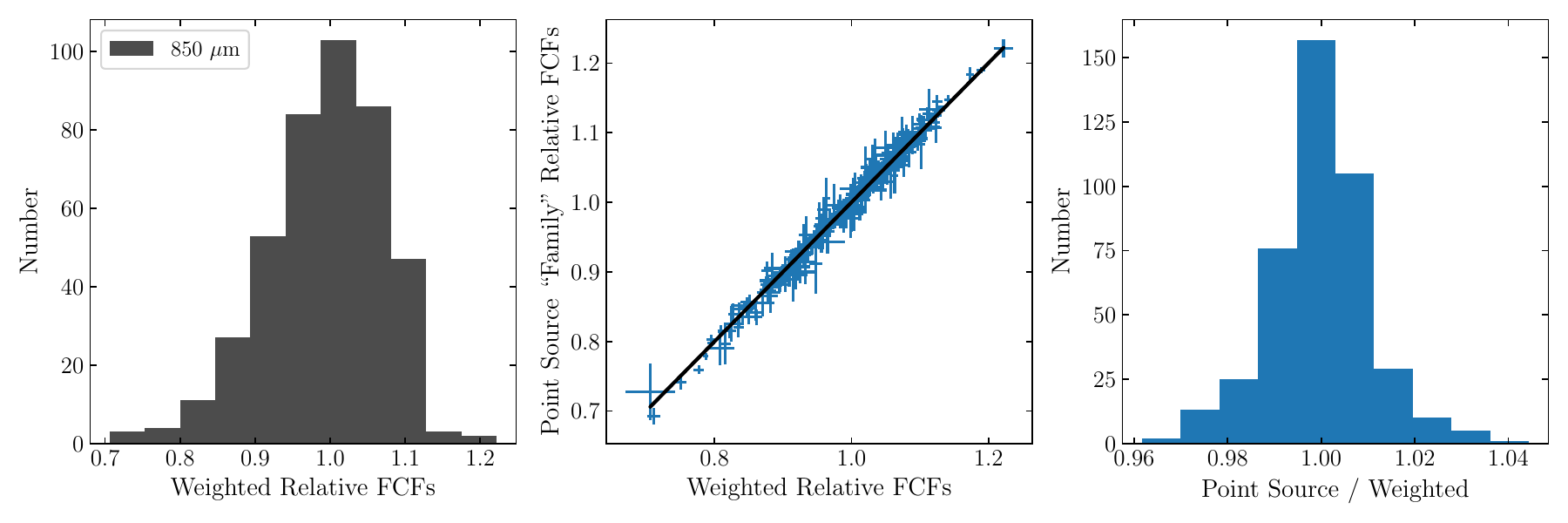}
\caption{All ``low''-mass regions, 850 $\mu$m. \textit{Left:} The Distribution of Relative FCFs computed using the ``weighted'' algorithm. \textit{Center:} Relative FCFs computed by the original Pipeline v1 (``Point Source'') as a function of the Relative FCF computed using the ``weighted'' algorithm (this work). A 1:1 line is overlaid. \textit{Right:} Distribution of the ``Point Source'' algorithm Relative FCFs divided by the ``weighted'' algorithm Relative FCFs, showing consistency between the two methods to within 2\%.}
\label{fig:FluxCalCompare850}
\end{figure*}

\begin{figure*}
\plotone{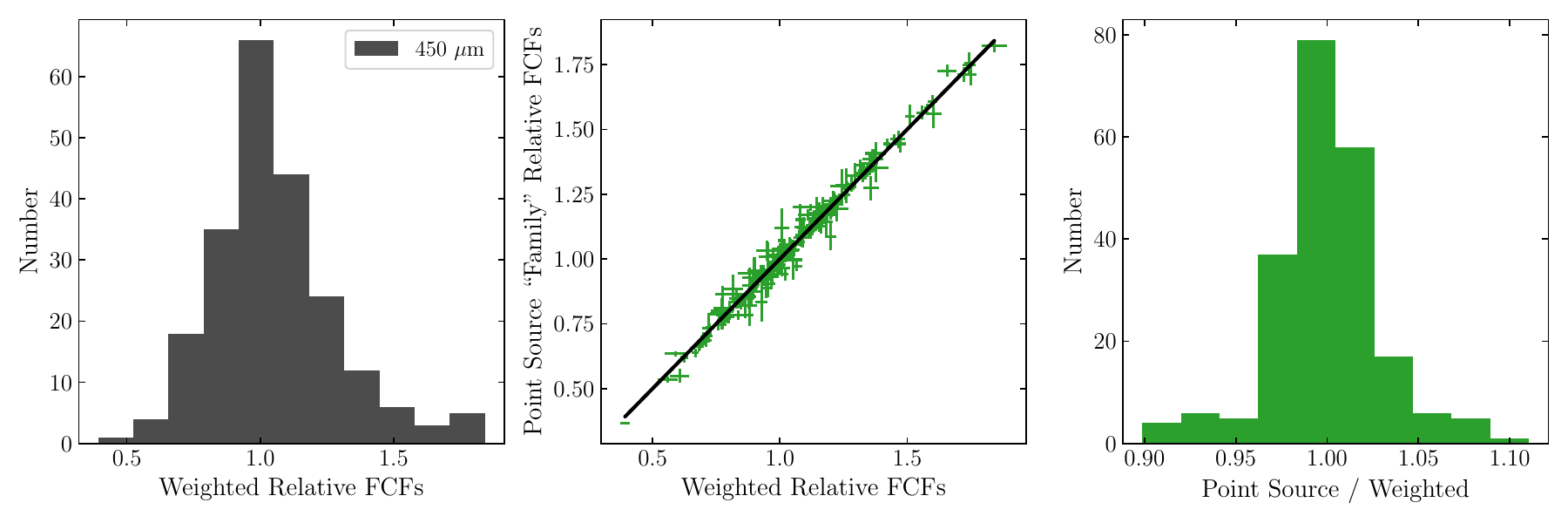}
\caption{Same as Figure \ref{fig:FluxCalCompare850}, but for the ``Good'' 450$\,\mu$m data (see Section \ref{subsec:450data}).}
\label{fig:FluxCalCompare450}
\end{figure*}

Figures \ref{fig:FluxCalCompare850} and 
\ref{fig:FluxCalCompare450} summarize the results of the full
relative flux calibration procedure. The left panels show the
distribution of all $\sigma_{\mathrm{FCF}}$ values derived for the eight Gould
Belt regions using the iterative, weighted algorithm 
described in Section \ref{sec:weightedsourcecal}. The 
450$\,\mu$m data included was observed in ``good-weather'' 
conditions and had a $\sigma_{\mathrm{FCF}}$ of $\leq5\%$, 
as described in Section \ref{subsec:450data}. The inherent 
telescope flux uncertainty can be estimated from the width of
each distribution's main peak. The width of the 850$\,\mu$m 
peak is 8-10\%, while the width of the 450$\,\mu$m peak is 
15-20\%, in good agreement with the uncertainties derived via
analyses of JCMT calibrator observations 
\citep{dempsey2013,mairs2021}. 

The center and right panels of Figures \ref{fig:FluxCalCompare850} and 
\ref{fig:FluxCalCompare450}
show the correlation between the \textit{Pipeline v2} flux calibration
algorithm with the \textit{Point Source method}.\footnote{A robust comparison between methods cannot be performed 
specifically for the IC\,348 region, however, because the 
\textit{Point Source method} relies on the the flux 
measurements of only two sources and one is near enough the 
minimum brightness threshold that the method can only produce
robust results for three individual observations.} 
The 850$\,\mu$m $R_{\mathrm{FCF}}$ values for these two methods agree to better than 3\% 
while the 450$\,\mu$m $\sigma_{\mathrm{FCF}}$ values agree to better than 7\%.
Furthermore, Table \ref{tab:GB-obs-summary} in Appendix 
\ref{app:GB-summary} allows for the determination of typical conversion
uncertainties at \LongNS, $\sigma_{\mathrm{FCF}} = 0.01$ or 1\%, and \ShortNS, $ 0.035 < \sigma_{\mathrm{FCF}} < 0.05$ or 3.5 to 5\%. These values can be used along with Equation \ref{eq:SDfid} when estimating the expected standard deviation of any particular source.


\begin{deluxetable*}{ccccccc}
\tablecaption{Summary of Gould Belt Region co-add images released.}
\label{tab:CoaddGB}
\tablecolumns{9}
\tablewidth{0pt}
\tablehead{
\colhead{Region} &
\colhead{R.A.} &
\colhead{Dec} &
\colhead{RMS$_{450}$} &
\colhead{RMS$_{850}$} &
\colhead{N$_{450}$}  &
\colhead{N$_{850}$} \\
\colhead{} &
\colhead{(J2000)} &
\colhead{(J2000)} &
\colhead{(mJy\,bm$^{-1}$)} &
\colhead{(mJy\,bm$^{-1}$)} &
\colhead{}  &
\colhead{}
}
\startdata
Perseus--IC348    & 03:44:18 & +32:04:59 & 31 & 1.5 & 25 & 50  \\
Perseus--NGC 1333 & 03:28:54 & +31:16:52 & 26 & 1.5 & 26 & 52  \\
Orion A--OMC 2/3  & 05:35:31 & -05:00:38 & 24 & 1.4 & 34 & 52  \\
Orion B--NGC 2024 & 05:41:41 & -01:53:51 & 28 & 1.5 & 36 & 49  \\
Orion B--NGC 2068 & 05:46:13 & -00:06:05 & 24 & 1.3 & 39 & 61  \\
Ophiuchus         & 16:27:05 & -24:32:37 & 22 & 1.8 & 19 & 37  \\
Serpens Main      & 18:29:49 & +01:15:20 & 24 & 1.2 & 48 & 79  \\
Serpens South     & 18:30:02 & -02:02:48 & 24 & 1.3 & 37 & 62  \\
\enddata
\end{deluxetable*}

\begin{figure*}
\centering
\includegraphics[width=0.80\textwidth]{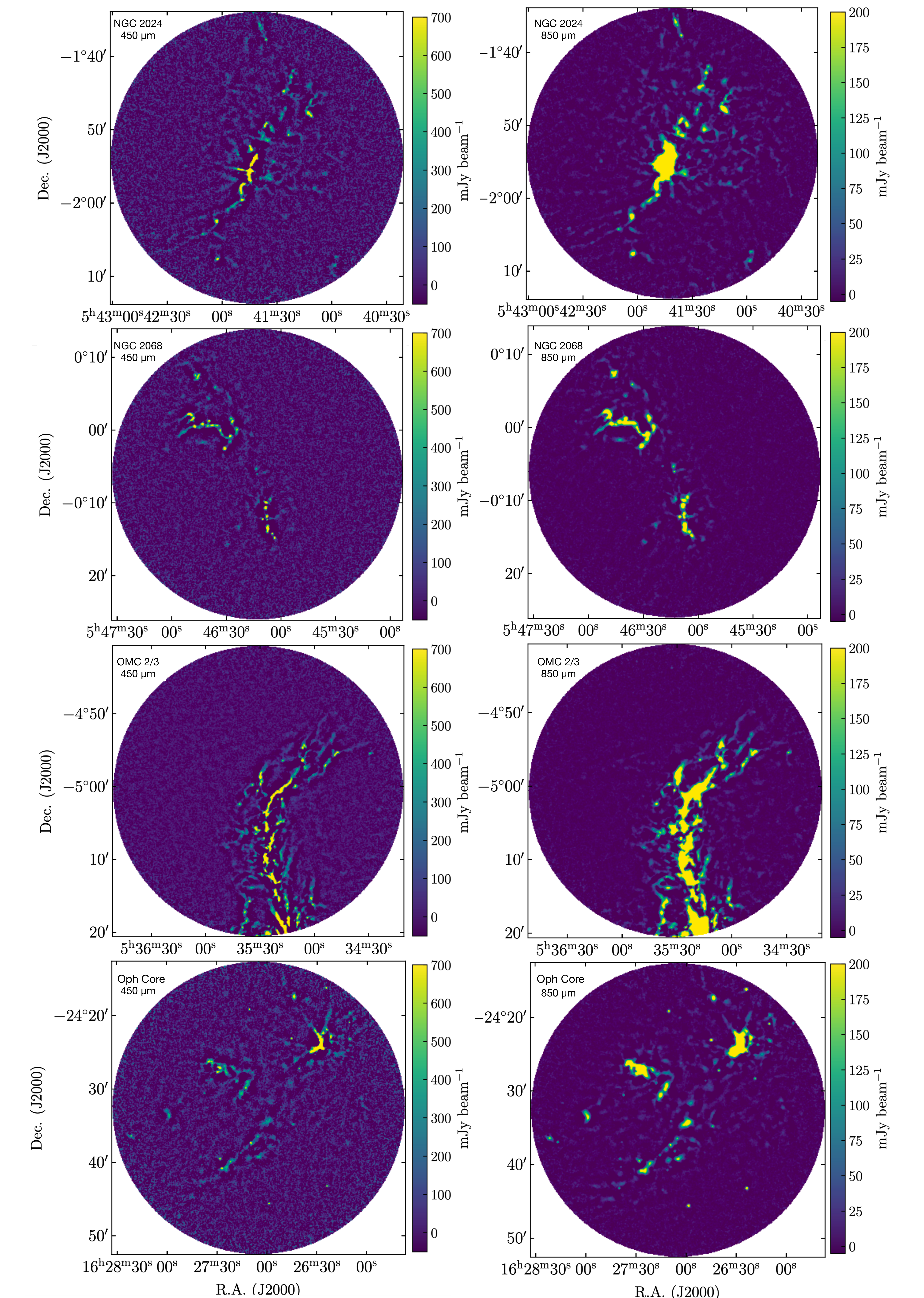}
\caption{Co-added images of NGC\,2024, NGC\,2068, OMC\,2/3 and Ophiuchus Core (top to bottom, respectively) at both 450 (left) and 850$\,\mu$m (right). Observations through February 2022 are included. At 450$\,\mu$m, only ``usable'' data as defined in Section \ref{subsec:450data} are included.}
\label{fig:CoaddsOrionOph}
\end{figure*}
\begin{figure*}
\centering
\includegraphics[width=0.80\textwidth]{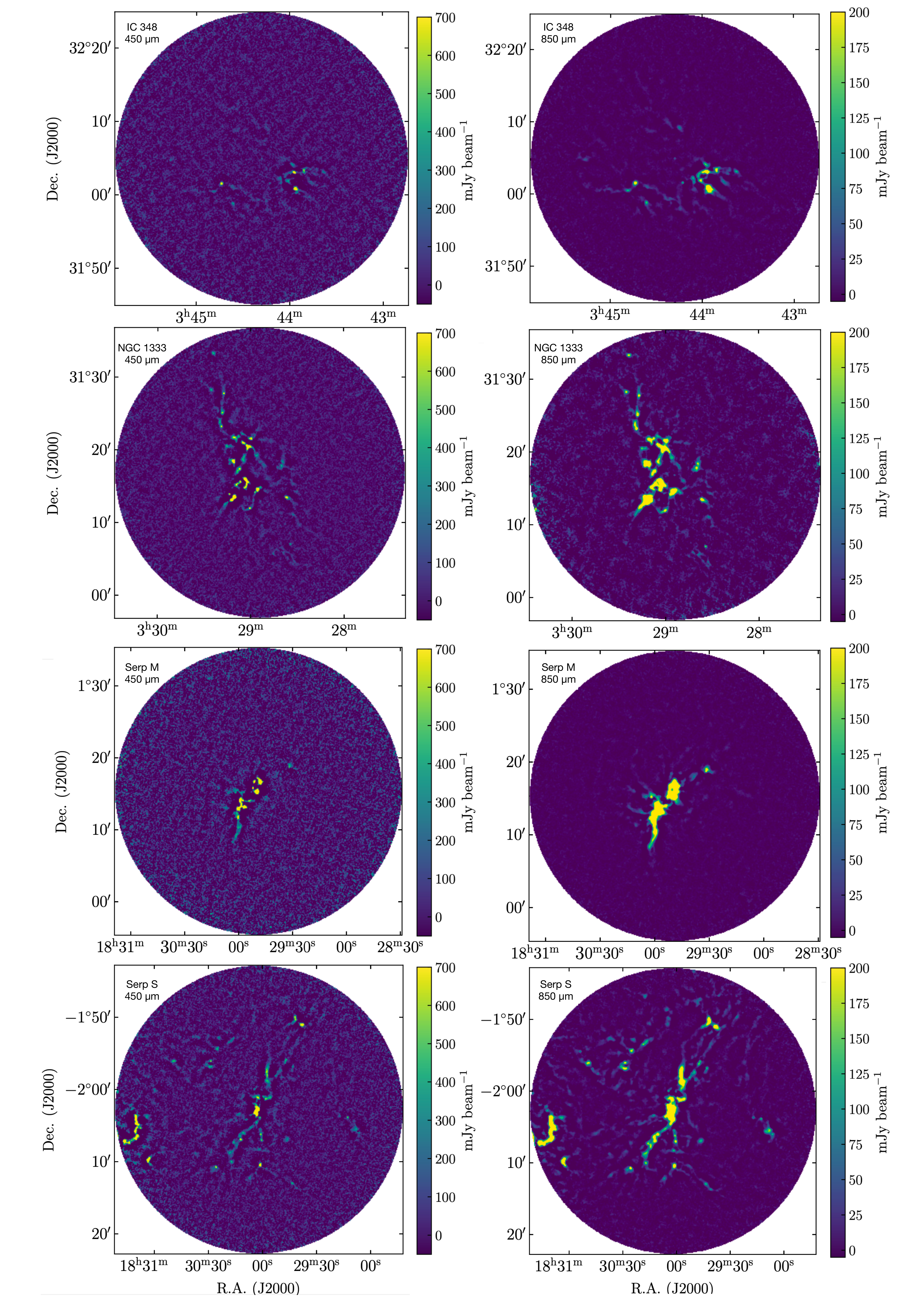}
\caption{Co-added images of IC\,348, NGC\,1333, Serpens Main, and Serpens South (top to bottom, respectively) at both 450 (left) and 850$\,\mu$m (right). Observations through February 2022 are included. At 450$\,\mu$m, only ``usable'' data as defined in Section \ref{subsec:450data} are included.}
\label{fig:CoaddsPerseusSerp}
\end{figure*}

\section{Co-added Images}
\label{sec:coadds}

The JCMT Transient Survey's consistent, repeated observing strategy from 2015 December through 2022 February has led to the deepest 450 and 850\,$\mu$m maps to-date of eight 
$\sim30\arcmin$-diameter Gould Belt fields. Figures 
\ref{fig:CoaddsOrionOph} and \ref{fig:CoaddsPerseusSerp} show
each cross-correlation-aligned relative flux calibrated co-add over the same, 
wavelength-consistent color scales and Table 
\ref{tab:CoaddGB} summarizes the number of epochs included in
each image along with the background RMS. At \ShortNS, only 
the ``usable'' maps, as defined in Section 
\ref{subsec:450data} were included. 

The images released in this work should not be used to 
analyze large-scale ($>3\arcmin$), flocculent gas and dust 
structures. The data reduction parameters specifically chosen
for this work were optimized to accurately recover compact, 
peak fluxes over a large dynamic range at the expense of 
reconstructing faint, less dense regions. In fact, while 
SCUBA-2's combined PONG1800 mapping strategy and data 
reduction procedure allow for some degree of separation of 
atmospheric and astronomical signal on scales up to 
$\sim10\arcmin$, ground and airborne-based submillimetre
bolometer data always suffer from spatial filtering on scales
larger than the characteristic detector size, leading to flux
loss on the edges of large structures (see 
\citealt{chapin2013}, \citealt{mairs2015}, \citealt{mairs2017} and 
\citealt{mairs2017GBSTrans} for more details). Co-added images are available here: \url{https://www.canfar.net/citation/landing?doi=23.0009}.

\section{Results}
\label{sec:results}

In this section, for consistent comparisons with previous results, we use the methods outlined by \citet{yhlee2021} to redetermine the long-term, secular variables separately at \LongS and \ShortS in order to test the updated calibrations. The secular variability is based on both linear and sinusoidal fits to the time-domain observations \citep[see][]{lomb1976, scargle89, vanderplas18}, with the best-fit False Alarm Probability (FAP) for interesting sources presented in Table \ref{tab:FAP}. Additional information about specific robust secular variables can be found in the appendix to \citet{yhlee2021}. 

We start by comparing the variables recovered by the new \LongS calibration (\textit{Pipeline v2}) against those recovered by the original scheme (\textit{Pipeline v1}), taking the same 4\,year time window as used by \citet{yhlee2021} (Section \ref{sec:results850_4yr}). We next consider the advantage of a longer time window when searching for secular valuables, utilizing 6\,years of monitoring at \LongS (Section \ref{sec:results850_6yr}). Then, for the first time, we look for evidence of variability at \ShortNS, using all 6\,years of monitoring (section \ref{sec:results450_850}) and briefly investigate the combined results.

\subsection{Robustness of \texorpdfstring{850\,$\mu$m}{850 micron} Variables Over 4 Years}
\label{sec:results850_4yr}

\begin{figure*}
\plotone{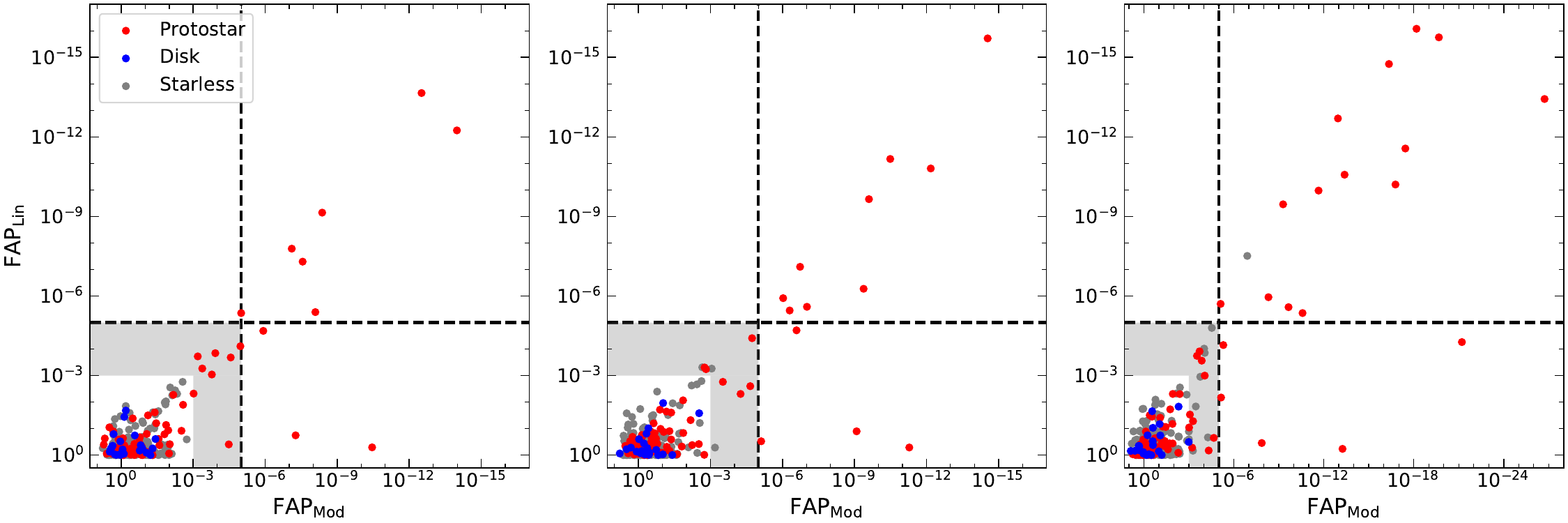}
\caption{Scatter plot of sinusoidal (FAP$_{\mathrm{Mod}}$) and linear (FAP$_{\mathrm{Lin}}$) false alarm probability at 850 microns. Left: Four years of data using the original alignment and reduction pipeline as performed by \citet{yhlee2021}. Middle: Four years of data reduced with the new calibration pipeline introduced in this paper. Right: Six years of data reduced with the new calibration pipeline. Note that the x-axis extends to much smaller false alarm values in the Right panel compared with the Left and Centre panels.}
\label{fig:850_FAP}
\end{figure*}

Utilizing the original calibration methods \citep[\textit{Pipeline v1},][]{mairs2017} and the first four years of the JCMT Transient Survey observations, through January 2020, \citet{yhlee2021} found 18 secular variables. Based on best-fit Lomb-Scargle Periodogram \citep{lomb1976, scargle89, vanderplas18} derived periods $P$, 2 of these are periodic ($P < 4$\,yr) and 11 are curved ($4$\,yr$\ < P < 15$\,yr), while 5 are best fit as linear-slope sources. In all cases, the FAP was required to be $< 10^{-3}$ (Figure \ref{fig:850_FAP}, Left Panel). These secular variables were all protostars, no prestellar or disk sources were observed to vary. 

Using these methods, every lightcurve has a best-fit sinusoid, regardless if it is a periodic, curved, linear, or even a non-variable source. The FAP value for non-variable sources, however, will be high, indicating a poor fit. Furthermore, as these period fits assume an underlying sinusoid the derived values should be taken as representative of the quantitative nature, rather than specific. For timescales shorter than the observing window the periods reasonable measure of episodic timescale whereas for longer derived periods the uncertainties are necessarily much larger. Given the $<10$-year timescale of these observations, there is insufficient data to distinguish linearly varying sources from sources with sinusoidal variability over long periods, since sinusoids can be regarded as a straight line near its zeros when the period is long. Variables with periods longer than 20 years are, therefore, classified as 'linear', but also assigned a derived period (see \citealt{park2021,yhlee2021} for further discussion). Further observations over the coming years may refine these fits and allow for differentiation in the future.

With our new calibrations (\textit{Pipeline v2}) and using the same four years of data and FAP threshold (Figure \ref{fig:850_FAP}, Middle Panel), we find 22 secular variables; 3 periodic, 14 curved, and 5 linear sources. Modifying slightly the classification from \citet{yhlee2021}, we consider those sources with FAP $< 10^{-5}$ to be robust secular variables, and those with $10^{-5} <$ FAP $< 10^{-3}$ to be candidate secular variables. With this definition, \citet{yhlee2021} recovered 10 robust and 8 candidates  while the new calibrations recover 13 robust and 9 candidates. Robust fit parameters are presented in Table \ref{tab:4yr_850}.

\begin{longtable*}[b]{|c|c|c|c|c|c|c|c|c|c|}
\caption{Log(FAP) for Candidate (yellow) and Robust (green) Variables.} 
\label{tab:FAP} \\

\hline \multicolumn{1}{|c|}{\textbf{ID 850\,$\mu$m}} & \multicolumn{1}{c|}{\textbf{ID 450\,$\mu$m}} & \multicolumn{1}{c|}{\textbf{850\,v1}} & \multicolumn{1}{c|}{\textbf{850\,v1}} & \multicolumn{1}{c|}{\textbf{850\,v2}} & \multicolumn{1}{c|}{\textbf{850\,v2}} & \multicolumn{1}{c|}{\textbf{850\,v2}} & \multicolumn{1}{c|}{\textbf{850\,v2}} & \multicolumn{1}{c|}{\textbf{450\,v2}} & \multicolumn{1}{c|}{\textbf{450\,v2}} \\ 
\multicolumn{1}{|c|}{} & \multicolumn{1}{c|}{} & \multicolumn{1}{c|}{4\,yr} & \multicolumn{1}{c|}{4\,yr} & \multicolumn{1}{c|}{4\,yr} & \multicolumn{1}{c|}{4\,yr} & \multicolumn{1}{c|}{6\,yr} & \multicolumn{1}{c|}{6\,yr} & \multicolumn{1}{c|}{6\,yr} & \multicolumn{1}{c|}{6\,yr}\\
\multicolumn{1}{|c|}{} & \multicolumn{1}{c|}{} & \multicolumn{1}{c|}{Linear} & \multicolumn{1}{c|}{Sinusoid} & \multicolumn{1}{c|}{Linear} & \multicolumn{1}{c|}{Sinusoid} & \multicolumn{1}{c|}{Linear} & \multicolumn{1}{c|}{Sinusoid} & \multicolumn{1}{c|}{Linear} & \multicolumn{1}{c|}{Sinusoid}\\
\hline 
\endfirsthead

\multicolumn{10}{c}%
{{\bfseries \tablename\ \thetable{} -- continued from previous page}} \\
\hline \multicolumn{1}{|c|}{\textbf{ID 850\,$\mu$m}} & \multicolumn{1}{c|}{\textbf{ID 450\,$\mu$m}} & \multicolumn{1}{c|}{\textbf{850 v1}} & \multicolumn{1}{c|}{\textbf{850 v1}} & \multicolumn{1}{c|}{\textbf{850 v2}} & \multicolumn{1}{c|}{\textbf{850 v2}} & \multicolumn{1}{c|}{\textbf{850 v2}} & \multicolumn{1}{c|}{\textbf{850 v2}} & \multicolumn{1}{c|}{\textbf{450 v2}} & \multicolumn{1}{c|}{\textbf{450 v2}} \\ 
\multicolumn{1}{|c|}{} & \multicolumn{1}{c|}{} & \multicolumn{1}{c|}{4\,yr} & \multicolumn{1}{c|}{4\,yr} & \multicolumn{1}{c|}{4\,yr} & \multicolumn{1}{c|}{4\,yr} & \multicolumn{1}{c|}{6\,yr} & \multicolumn{1}{c|}{6\,yr} & \multicolumn{1}{c|}{6\,yr} & \multicolumn{1}{c|}{6\,yr}\\
\multicolumn{1}{|c|}{} & \multicolumn{1}{c|}{} & \multicolumn{1}{c|}{Linear} & \multicolumn{1}{c|}{Sinusoid} & \multicolumn{1}{c|}{Linear} & \multicolumn{1}{c|}{Sinusoid} & \multicolumn{1}{c|}{Linear} & \multicolumn{1}{c|}{Sinusoid} & \multicolumn{1}{c|}{Linear} & \multicolumn{1}{c|}{Sinusoid}\\
\hline 
\endhead

\hline \multicolumn{3}{|r|}{{Continued on next page}} \\ \hline
\endfoot
\hline \hline
\endlastfoot
J034356.5+320050 & J034356.6+320049 & \cellcolor{yellow!25}-3.267 & \cellcolor{yellow!25}-3.375 & \cellcolor{white!25}-0.891 & \cellcolor{white!25}-0.51 & \cellcolor{white!25}-0.442 & \cellcolor{white!25}-1.719 & \cellcolor{white!25}-0.189 & \cellcolor{white!25}-0.676\\
J034357.0+320305 & J034356.9+320305 & \cellcolor{green!25}-7.297 & \cellcolor{green!25}-7.557 & \cellcolor{green!25}-7.102 & \cellcolor{green!25}-6.735 & \cellcolor{green!25}-5.58 & \cellcolor{green!25}-9.648 & \cellcolor{white!25}-2.06 & \cellcolor{white!25}-2.047\\
J032910.4+311331 & J032910.4+311330 & \cellcolor{yellow!25}-3.04 & \cellcolor{yellow!25}-3.775 & \cellcolor{yellow!25}-4.706 & \cellcolor{green!25}-6.586 & \cellcolor{green!25}-5.358 & \cellcolor{green!25}-10.583 & \cellcolor{white!25}-1.2 & \cellcolor{white!25}0.056\\
J032912.0+311307 & J032911.9+311308 & \cellcolor{white!25}-0.621 & \cellcolor{white!25}-0.752 & \cellcolor{white!25}-0.379 & \cellcolor{white!25}-2.239 & \cellcolor{white!25}-2.998 & \cellcolor{yellow!25}-4.058 & \cellcolor{white!25}-0.087 & \cellcolor{white!25}-0.309\\
J032903.4+311558 &--& \cellcolor{green!25}-7.786 & \cellcolor{green!25}-7.104 & \cellcolor{green!25}-11.169 & \cellcolor{green!25}-10.491 & \cellcolor{green!25}-16.079 & \cellcolor{green!25}-18.184 & \cellcolor{white!25}-- & \cellcolor{white!25}--\\
J032906.6+311537 &--& \cellcolor{white!25}-2.763 & \cellcolor{white!25}-2.557 & \cellcolor{white!25}-1.876 & \cellcolor{white!25}-1.722 & \cellcolor{yellow!25}-4.797 & \cellcolor{yellow!25}-4.53 & \cellcolor{white!25}-- & \cellcolor{white!25}--\\
J032903.8+311449 &--& \cellcolor{green!25}-9.145 & \cellcolor{green!25}-8.372 & \cellcolor{green!25}-9.653 & \cellcolor{green!25}-9.604 & \cellcolor{green!25}-14.75 & \cellcolor{green!25}-16.348 & \cellcolor{white!25}-- & \cellcolor{white!25}--\\
J032911.1+311828 & J032911.2+311830 & \cellcolor{white!25}-0.411 & \cellcolor{white!25}-2.023 & \cellcolor{white!25}-2.6 & \cellcolor{yellow!25}-4.659 & \cellcolor{white!25}-0.651 & \cellcolor{yellow!25}-4.663 & \cellcolor{white!25}-0.002 & \cellcolor{white!25}-0.044\\
J032901.5+312028 & J032901.3+312028 & \cellcolor{white!25}-0.082 & \cellcolor{white!25}-0.273 & \cellcolor{white!25}-0.454 & \cellcolor{white!25}0.107 & \cellcolor{yellow!25}-3.744 & \cellcolor{yellow!25}-3.539 & \cellcolor{white!25}-0.91 & \cellcolor{white!25}-0.319\\
J032907.8+312155 &--& \cellcolor{white!25}-0.019 & \cellcolor{white!25}-0.399 & \cellcolor{white!25}-0.268 & \cellcolor{white!25}-0.363 & \cellcolor{yellow!25}-3.563 & \cellcolor{yellow!25}-3.856 & \cellcolor{white!25}-- & \cellcolor{white!25}--\\
J032836.9+311328 & J032837.2+311330 & \cellcolor{white!25}-0.047 & \cellcolor{white!25}0.364 & \cellcolor{white!25}-0.622 & \cellcolor{white!25}-0.734 & \cellcolor{white!25}-1.076 & \cellcolor{white!25}-1.26 & \cellcolor{white!25}-2.818 & \cellcolor{yellow!25}-3.041\\
J054202.8-020230 & J054202.9-020233 & \cellcolor{white!25}-0.139 & \cellcolor{white!25}-0.602 & \cellcolor{white!25}-0.298 & \cellcolor{white!25}-0.095 & \cellcolor{yellow!25}-4.018 & \cellcolor{yellow!25}-4.017 & \cellcolor{white!25}-0.111 & \cellcolor{white!25}0.255\\
J054124.2-014433 &--& \cellcolor{white!25}-2.551 & \cellcolor{white!25}-2.039 & \cellcolor{yellow!25}-3.311 & \cellcolor{white!25}-2.658 & \cellcolor{white!25}-1.835 & \cellcolor{yellow!25}-3.46 & \cellcolor{white!25}-- & \cellcolor{white!25}--\\
J054608.4-001041 & J054608.3-001041 & \cellcolor{white!25}-1.898 & \cellcolor{white!25}-2.576 & \cellcolor{yellow!25}-3.306 & \cellcolor{white!25}-2.755 & \cellcolor{green!25}-12.697 & \cellcolor{green!25}-12.943 & \cellcolor{yellow!25}-3.263 & \cellcolor{white!25}-2.761\\
J054607.2-001332 & J054607.3-001329 & \cellcolor{green!25}-13.656 & \cellcolor{green!25}-12.512 & \cellcolor{green!25}-15.717 & \cellcolor{green!25}-14.549 & \cellcolor{yellow!25}-4.263 & \cellcolor{green!25}-21.214 & \cellcolor{white!25}-1.713 & \cellcolor{green!25}-11.477\\
J054603.6-001447 & J054603.5-001447 & \cellcolor{white!25}-2.32 & \cellcolor{yellow!25}-3.014 & \cellcolor{white!25}-2.307 & \cellcolor{yellow!25}-4.255 & \cellcolor{green!25}-11.564 & \cellcolor{green!25}-17.44 & \cellcolor{green!25}-6.417 & \cellcolor{green!25}-6.151\\
J054604.8-001417 &--& \cellcolor{white!25}-0.458 & \cellcolor{white!25}-0.029 & \cellcolor{white!25}-0.412 & \cellcolor{white!25}-0.159 & \cellcolor{yellow!25}-4.15 & \cellcolor{green!25}-5.297 & \cellcolor{white!25}-- & \cellcolor{white!25}--\\
J054631.0-000232 & J054630.9-000233 & \cellcolor{white!25}-0.746 & \cellcolor{green!25}-7.265 & \cellcolor{white!25}-0.898 & \cellcolor{green!25}-9.09 & \cellcolor{green!25}-13.431 & \cellcolor{green!25}-26.727 & \cellcolor{green!25}-7.476 & \cellcolor{green!25}-12.211\\
J054647.4+000028 & J054647.4+000027 & \cellcolor{yellow!25}-4.106 & \cellcolor{yellow!25}-4.964 & \cellcolor{white!25}-2.763 & \cellcolor{yellow!25}-3.518 & \cellcolor{yellow!25}-3.913 & \cellcolor{yellow!25}-3.711 & \cellcolor{white!25}-2.768 & \cellcolor{white!25}-2.555\\
J054637.8+000037 & J054637.8+000035 & \cellcolor{white!25}-0.074 & \cellcolor{white!25}-1.237 & \cellcolor{white!25}-0.264 & \cellcolor{white!25}-0.441 & \cellcolor{white!25}-0.282 & \cellcolor{yellow!25}-3.217 & \cellcolor{white!25}-0.949 & \cellcolor{white!25}-0.761\\
J054639.4+000113 & J054639.3+000113 & \cellcolor{white!25}-0.451 & \cellcolor{white!25}-0.767 & \cellcolor{white!25}-0.525 & \cellcolor{white!25}0.349 & \cellcolor{white!25}-1.281 & \cellcolor{yellow!25}-3.276 & \cellcolor{white!25}-0.082 & \cellcolor{white!25}0.049\\
J054645.0+000022 &--& \cellcolor{white!25}-0.007 & \cellcolor{white!25}-1.087 & \cellcolor{white!25}-0.008 & \cellcolor{white!25}-0.835 & \cellcolor{white!25}-0.166 & \cellcolor{yellow!25}-3.272 & \cellcolor{white!25}-- & \cellcolor{white!25}--\\
J054613.2-000602 &--& \cellcolor{green!25}-12.247 & \cellcolor{green!25}-13.986 & \cellcolor{green!25}-10.811 & \cellcolor{green!25}-12.18 & \cellcolor{green!25}-15.755 & \cellcolor{green!25}-19.678 & \cellcolor{white!25}-- & \cellcolor{white!25}--\\
J054636.2+000552 &--& \cellcolor{white!25}-0.677 & \cellcolor{white!25}-0.287 & \cellcolor{white!25}-0.573 & \cellcolor{white!25}-0.158 & \cellcolor{white!25}-0.598 & \cellcolor{yellow!25}-3.005 & \cellcolor{white!25}-- & \cellcolor{white!25}--\\
J053527.4-050929 & J053527.5-050930 & \cellcolor{white!25}-0.918 & \cellcolor{white!25}-2.509 & \cellcolor{white!25}-0.416 & \cellcolor{white!25}-2.518 & \cellcolor{white!25}-1.529 & \cellcolor{yellow!25}-3.069 & \cellcolor{white!25}-0.893 & \cellcolor{yellow!25}-3.325\\
J053521.6-051308 & J053521.4-051312 & \cellcolor{white!25}-0.118 & \cellcolor{white!25}0.077 & \cellcolor{white!25}-0.001 & \cellcolor{white!25}-0.076 & \cellcolor{white!25}-0.002 & \cellcolor{white!25}-0.366 & \cellcolor{white!25}-2.462 & \cellcolor{yellow!25}-3.453\\
J053524.4-050829 & J053524.3-050828 & \cellcolor{white!25}-1.046 & \cellcolor{white!25}0.496 & \cellcolor{white!25}-1.713 & \cellcolor{white!25}-0.896 & \cellcolor{white!25}-1.449 & \cellcolor{white!25}-0.54 & \cellcolor{white!25}-2.601 & \cellcolor{yellow!25}-3.506\\
J053523.4-051202 & J053523.2-051200 & \cellcolor{white!25}-0.001 & \cellcolor{white!25}-1.305 & \cellcolor{white!25}-0.012 & \cellcolor{white!25}-2.743 & \cellcolor{white!25}-1.062 & \cellcolor{white!25}-1.455 & \cellcolor{white!25}-2.879 & \cellcolor{yellow!25}-4.184\\
J053519.6-051529 & J053519.5-051532 & \cellcolor{white!25}-0.611 & \cellcolor{white!25}-1.557 & \cellcolor{white!25}-0.903 & \cellcolor{white!25}-1.293 & \cellcolor{green!25}-10.205 & \cellcolor{green!25}-16.78 & \cellcolor{white!25}-1.995 & \cellcolor{yellow!25}-3.655\\
J053523.4-050126 & J053523.4-050126 & \cellcolor{white!25}-1.611 & \cellcolor{white!25}-1.397 & \cellcolor{white!25}-2.066 & \cellcolor{white!25}-1.859 & \cellcolor{green!25}-5.703 & \cellcolor{green!25}-5.117 & \cellcolor{white!25}-2.644 & \cellcolor{yellow!25}-4.489\\
J053522.4-050111 & J053522.4-050110 & \cellcolor{white!25}-0.675 & \cellcolor{white!25}-0.773 & \cellcolor{white!25}-0.428 & \cellcolor{white!25}-0.786 & \cellcolor{white!25}-1.042 & \cellcolor{yellow!25}-3.015 & \cellcolor{white!25}-1.945 & \cellcolor{yellow!25}-3.789\\
J053518.0-050017 & J053518.0-050016 & \cellcolor{white!25}-1.615 & \cellcolor{white!25}-1.316 & \cellcolor{white!25}-0.837 & \cellcolor{white!25}-1.771 & \cellcolor{yellow!25}-3.856 & \cellcolor{yellow!25}-4.061 & \cellcolor{white!25}-1.381 & \cellcolor{yellow!25}-3.275\\
J053529.8-045944 &--& \cellcolor{green!25}-5.393 & \cellcolor{green!25}-8.088 & \cellcolor{green!25}-6.273 & \cellcolor{green!25}-9.381 & \cellcolor{green!25}-9.977 & \cellcolor{green!25}-11.652 & \cellcolor{white!25}-- & \cellcolor{white!25}--\\
J162626.8-242431 & J162626.6-242431 & \cellcolor{yellow!25}-3.723 & \cellcolor{yellow!25}-3.188 & \cellcolor{yellow!25}-4.41 & \cellcolor{yellow!25}-4.734 & \cellcolor{green!25}-9.459 & \cellcolor{green!25}-9.285 & \cellcolor{white!25}-1.785 & \cellcolor{white!25}-1.512\\
J182949.8+011520 & J182949.8+011520 & \cellcolor{yellow!25}-4.686 & \cellcolor{green!25}-5.918 & \cellcolor{green!25}-5.594 & \cellcolor{green!25}-7.02 & \cellcolor{green!25}-10.576 & \cellcolor{green!25}-13.385 & \cellcolor{white!25}-1.473 & \cellcolor{white!25}-0.641\\
J182948.2+011644 & J182948.1+011642 & \cellcolor{yellow!25}-3.845 & \cellcolor{yellow!25}-3.919 & \cellcolor{yellow!25}-3.242 & \cellcolor{white!25}-2.814 & \cellcolor{white!25}-0.456 & \cellcolor{green!25}-7.862 & \cellcolor{white!25}-0.995 & \cellcolor{yellow!25}-3.572\\
J182951.2+011638 & J182951.3+011638 & \cellcolor{white!25}-0.293 & \cellcolor{green!25}-10.452 & \cellcolor{white!25}-0.288 & \cellcolor{green!25}-11.285 & \cellcolor{white!25}-0.236 & \cellcolor{green!25}-13.248 & \cellcolor{white!25}-0.948 & \cellcolor{green!25}-7.249\\
J182952.0+011550 & J182952.1+011548 & \cellcolor{white!25}-0.405 & \cellcolor{yellow!25}-4.48 & \cellcolor{white!25}-0.522 & \cellcolor{green!25}-5.114 & \cellcolor{white!25}-0.173 & \cellcolor{yellow!25}-4.321 & \cellcolor{white!25}-0.013 & \cellcolor{white!25}0.261\\
J182947.6+011553 &--& \cellcolor{white!25}-1.102 & \cellcolor{white!25}-0.59 & \cellcolor{white!25}-1.738 & \cellcolor{white!25}-0.074 & \cellcolor{green!25}-7.513 & \cellcolor{green!25}-6.895 & \cellcolor{white!25}-- & \cellcolor{white!25}--\\
J183000.4+011144 & J183000.3+011142 & \cellcolor{white!25}-0.445 & \cellcolor{white!25}-1.123 & \cellcolor{white!25}-1.304 & \cellcolor{white!25}-0.586 & \cellcolor{white!25}-0.669 & \cellcolor{yellow!25}-4.268 & \cellcolor{white!25}-0.178 & \cellcolor{white!25}0.004\\
J182954.4+011359 &--& \cellcolor{white!25}-2.255 & \cellcolor{white!25}-2.123 & \cellcolor{white!25}-2.792 & \cellcolor{white!25}-2.626 & \cellcolor{white!25}-2.954 & \cellcolor{yellow!25}-3.758 & \cellcolor{white!25}-- & \cellcolor{white!25}--\\
J183004.0-020306 & J183004.0-020306 & \cellcolor{green!25}-5.357 & \cellcolor{yellow!25}-4.996 & \cellcolor{green!25}-5.919 & \cellcolor{green!25}-6.029 & \cellcolor{white!25}-2.172 & \cellcolor{green!25}-5.148 & \cellcolor{white!25}-0.056 & \cellcolor{white!25}-0.11\\
J183012.6-020627 &--& \cellcolor{white!25}-1.912 & \cellcolor{white!25}-1.842 & \cellcolor{yellow!25}-3.263 & \cellcolor{yellow!25}-3.051 & \cellcolor{white!25}-0.892 & \cellcolor{white!25}-2.998 & \cellcolor{white!25}-- & \cellcolor{white!25}--\\
J182937.8-015103 & J182937.6-015102 & \cellcolor{yellow!25}-3.683 & \cellcolor{yellow!25}-4.561 & \cellcolor{green!25}-5.455 & \cellcolor{green!25}-6.3 & \cellcolor{green!25}-5.956 & \cellcolor{green!25}-8.308 & \cellcolor{white!25}-0.093 & \cellcolor{white!25}0.835\\
J183103.4-020948 & J183103.5-020946 & \cellcolor{white!25}-0.013 & \cellcolor{white!25}-1.519 & \cellcolor{white!25}-0.284 & \cellcolor{yellow!25}-3.189 & \cellcolor{white!25}-0.05 & \cellcolor{white!25}-2.391 & \cellcolor{white!25}-0.001 & \cellcolor{white!25}-0.616\\
0 & J053520.0-051800 & \cellcolor{white!25}-- & \cellcolor{white!25}-- & \cellcolor{white!25}-- & \cellcolor{white!25}-- & \cellcolor{white!25}-- & \cellcolor{white!25}-- & \cellcolor{white!25}-2.621 & \cellcolor{yellow!25}-3.008
\end{longtable*}

Comparing the FAP values for two sets of detected secular variables (see Table \ref{tab:FAP}), we find that one source, J034356.5+320050 in Perseus (also known as IC\,348\,HH\,211),\footnote{In Tables 3 and 4 by \citet{yhlee2021} the source names for HH\,211 and IC348 MMS\,1 are inadvertently reversed.} 
classified as (candidate) periodic by \citet{yhlee2021}, is rejected using the new calibration. Within the \citet{yhlee2021} sample, HH\,211 had the shortest period, $\sim1\,$yr, and the second largest FAP$= 10^{-3.3}$. With the new calibration, the sinusoidal fitting finds the same best period; however, the false alarm increases significantly to FAP$= 10^{-0.5}$.  Of the other seven  candidate sources from \citet{yhlee2021}, four remain candidate variables using the new calibration while three are elevated to robust detections. Furthermore, all ten of the \citet{yhlee2021} sources with FAP $< 10^{-5}$ remain robust with the new calibration. Finally, we note that five additional sources are classified as candidate secular variables using the new calibration, of which three are un-associated with known young stellar objects. All three of these sources, however, have FAPs within a factor of two of the cut-off threshold.

\begin{figure*}
\plottwo{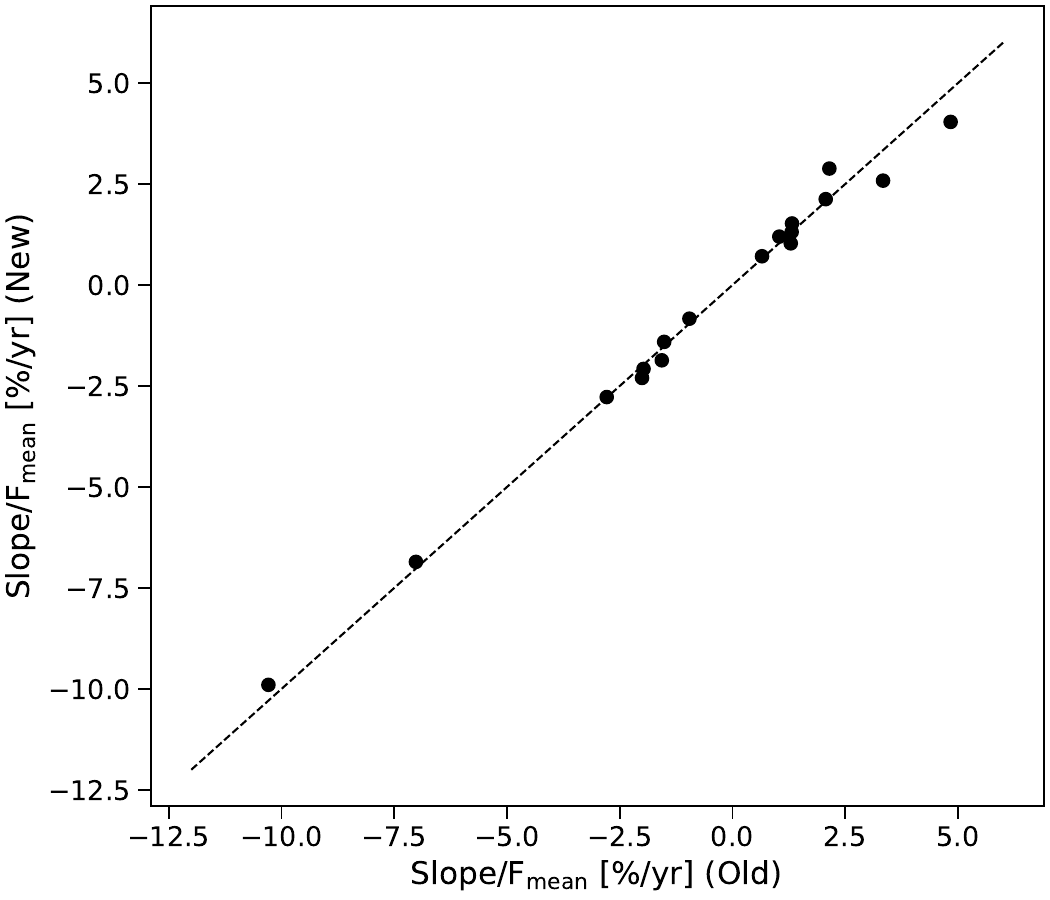}{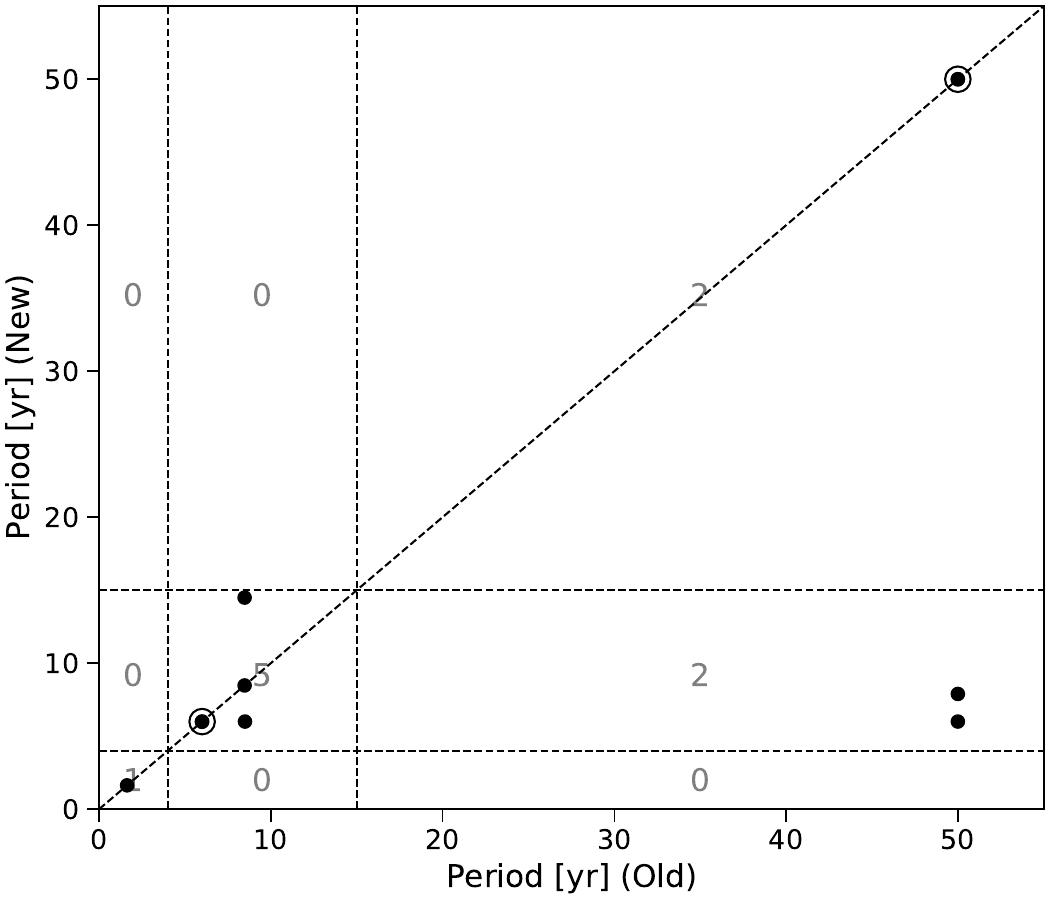}
\caption{Comparison of best-fit linear slope (left) and period (right) for old and new calibrations at 850 microns. For the slope comparison, all sources with with candidate or robust secular fits are included. For the period comparison, only those sources with robust periodic fits are included, and the dashed horizontal and vertical lines separate the plot into Periodic, Curved, and Linear regions. Two sources are overlapped in the period plot, so each is annotated with a bullseye.}
\label{fig:850_comp}
\end{figure*}

\begin{figure*}
\plotone{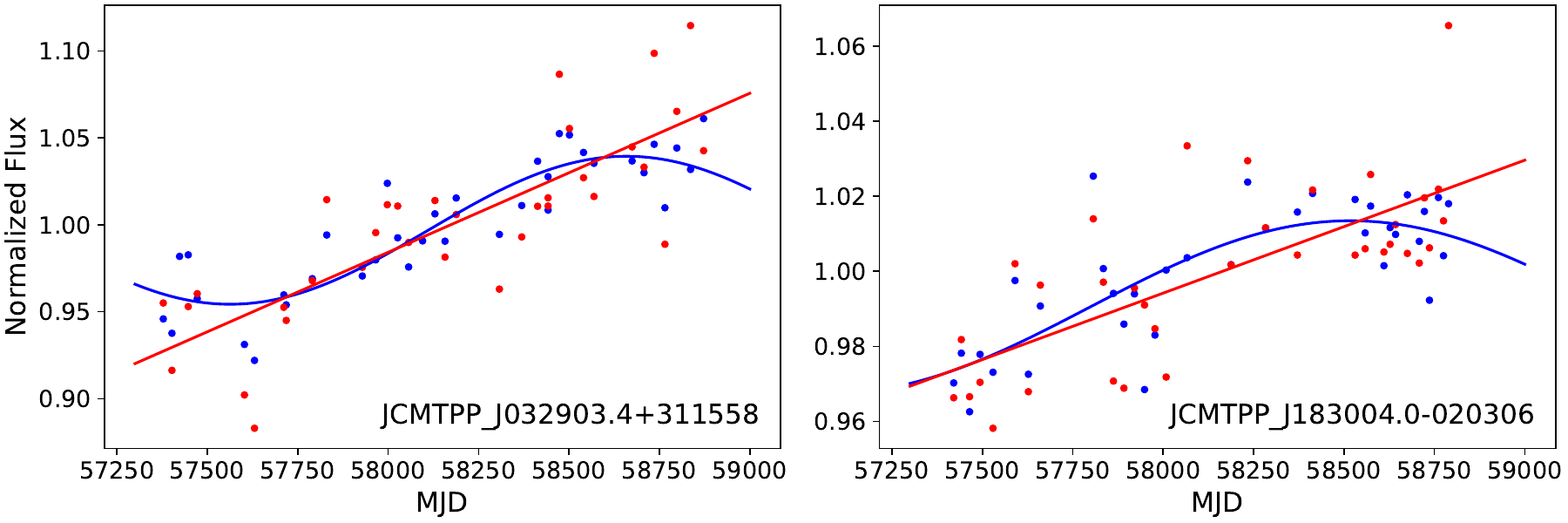}
\caption{Lightcurves of two sources that changed variable type, from linear (red - Pipeline v1) to  curved (blue - pipeline v2) after the improved calibration.}
\label{fig:cal_comparison}
\end{figure*}

Comparing our results between both of the calibration methods also provides a test on whether the calibration method affects the variability measurements and best-fit parameters.
The left panel of Figure \ref{fig:850_comp} plots the derived lightcurve linear slopes for the sources that were classified as variables, candidate and robust, for both calibration methods. The slopes are consistent between the calibrations, as expected. There are significant changes for a few of the fitted periods, however. The right panel of Figure \ref{fig:850_comp} shows the comparison of the best-fit periods for those ten sources with robust periodogram FAPs using the old calibration. Eight out of ten sources maintain their secular type (Periodic/Curved/Linear), though these sources often recover somewhat different best-fit periods between the calibrations. The other two sources switch between Linear and Curved type. Figure \ref{fig:cal_comparison} plots the old (red) and new (blue) calibrated measurements and best fits and show that the change to Curved type is subtle, depending specifically on the calibration of the early and late epochs. 

\begin{deluxetable*}{llccccc}
\tablecaption{Physical Properties of Robust Secular Variables at 850 Microns over 4 Years\label{tab:4yr_850}}
\small
\tablehead{
 \colhead{ID} & 
 \colhead{Known Name} & 
 \colhead{$\mathrm{\Bar{F}}$}  & 
 \colhead{Group} &
 \colhead{Period} & 
 \colhead{A/$\mathrm{\Bar{F}}$} & 
  \colhead{Slope/$\mathrm{\Bar{F}}$} \\
& & [Jy bm$^{-1}$]& & [yr]& [$\%$ yr$^{-1}$] & [$\%$] 
}
\startdata
J034357.0+320305& IC\,348 MMS\,1         & 1.3 & C& 6.0&	5.0 & 2.9 \\		
J032910.4+311331& NGC\,1333 IRAS\,4A     & 9.5 & C& 4.6& 4.3 &     \\				
J032903.4+311558& NGC\,1333 VLA\,3       & 3.2 & C& 6.0& 4.3 & 2.6 \\								
J032903.8+311449& NGC\,1333 West\,40     & 0.50& C& 6.0& 6.8 & 4.0 \\								
J054607.2-001332& NGC \,2068 HOPS\,358   & 1.3 & L& 50.&     &-6.9 \\	
J054631.0-000232& NGC \,2068 HOPS\,373   & 1.3 & C& 14.&	36. &	  \\		
J054613.2-000602& NGC \,2068 V\,1647\,Ori& 0.26& C& 6.0&  17.&-9.9 \\					
J053529.8-045944& OMC\,2/3 HOPS\,383     & 0.56& C& 8.5& 7.9 &-2.8 \\						
J182949.8+011520& Serpens Main SMM\,1    & 7.2 & L& 50.&     & 2.1 \\								
J182951.2+011638& Serpens Main EC\,53    & 1.2 & P& 1.6& 11. &     \\
J182952.0+011550& Serpens Main SMM\,10   & 0.85& C& 7.8& 7.8 &     \\								
J183004.0-020306& Serpens South CARMA\,7 & 4.8 & C& 7.9& 2.3  & 1.0 \\
J182937.8-015103& IRAS 18270-0153        & 0.63& L& 50.&	  &-2.3 
\enddata
\end{deluxetable*}
\begin{deluxetable*}{llccccc}
\tablecaption{Physical Properties of Robust Secular Variables at 850 Microns over 6 Years\label{tab:6yr_850}}
\small
\tablehead{
 \colhead{ID} & 
 \colhead{Common Name} & 
 \colhead{$\mathrm{\Bar{F}}$}  & 
 \colhead{Group} &
 \colhead{Period} & 
 \colhead{A/$\mathrm{\Bar{F}}$} & 
  \colhead{Slope/$\mathrm{\Bar{F}}$} \\
& & [Jy bm$^{-1}$]& & [yr]& [$\%$ yr$^{-1}$] & [$\%$] 
}
\startdata
J034357.0+320305& IC\,348 MMS\,1         &1.3 &C& 12.& 6.6&  1.5 \\
J032910.4+311331& NGC\,1333 IRAS\,4A     &9.3 &C& 6.7& 4.8& -1.4 \\
J032903.4+311558& NGC\,1333 VLA\,3       &3.2 &C& 12.& 5.3&  1.8 \\
J032903.8+311449& NGC\,1333 West\,40     &0.51&C& 8.5& 7.5&  3.0 \\
J054608.4-001041& NGC \,2068 LBS\,23\,SM &2.7 &L& 50.&    &  1.0 \\
J054607.2-001332& NGC \,2068 HOPS\,358   &1.3 &C& 8.5& 13.&      \\
J054603.6-001447& NGC \,2068 HOPS\,315   &0.56&L& 50.&    &  2.3 \\
J054604.8-001417& IRAS 05435-0015        &0.32&L& 50.&	  &      \\
J054631.0-000232& NGC \,2068 HOPS\,373   &1.4 &C& 8.5& 13.&  4.1 \\
J054613.2-000602& NGC \,2068 V\,1647\,Ori&0.24&C& 12.& 22.& -7.0 \\
J053519.6-051529& OMC\,2/3 HOPS\,56      &1.33&L& 50.&    &  2.5 \\
J053523.4-050126& OMC\,2/3 HOPS\,87      &5.8 &C& 12.& 2.1&  0.8 \\
J053529.8-045944& OMC\,2/3 HOPS\,383     &3.9 &C& 50.&    & -1.9 \\
J162626.8-242431& VLA 1623-243           &3.9 &C& 8.5& 4.1& -1.7 \\
J182949.8+011520& Serpens Main SMM\,1    &7.3 &L& 50.&    &  1.4 \\
J182948.2+011644&Serpens Main SH\,2-68\,N&2.1 &P& 5.4& 1.7&      \\
J182951.2+011638& Serpens Main EC\,53    &1.2 &P& 1.5& 9.5&      \\
J182947.6+011553& [see Section \ref{sec:results850_6yr}]          &0.41&C& 12.& 4.9&  2.0 \\
J183004.0-020306& Serpens South CARMA\,7 &4.8 &L& 50.&    & 0.3 \\
J182937.8-015103& IRAS 18270-0153        &0.63 &L& 50.&    & -1.3 
\enddata
\end{deluxetable*}

\subsection{Recovered \texorpdfstring{850\,$\mu$m}{850 micron} Variables at 6 yrs} 
\label{sec:results850_6yr}

Increasing the observing window from 4 years to 6 years, from the beginning of the survey through 2022 February, we update the criteria for secular variability accordingly. We categorize the variables by their best-fit periods with: periodic ($< 6$\,yr), curved ($6-20$\,yr), and Linear ($>20$\,yr). We now recover 38 secular variables, more than double the number obtained over 4 years by \citet{yhlee2021}. Of these sources, 20 are robust and 18 are candidate variables (Table \ref{tab:FAP}). Of the robust sources, 8 are linear, 10 are curved, and 2 are periodic (Table \ref{tab:6yr_850}). All robust sources are known protostars except the brightening curved source J182947.6+011553 in Serpens Main which is further discussed below. 

Of the candidate sources, 5 are linear, 6 are curved, and 7 are periodic. Half of the candidate periodic sources have very short periods, less than a year, and FAPs close to the cut-off value. Another 7 of the candidate sources are not known to be associated with YSOs. We suspect that systematic issues, potentially associated with the yearly weather patterns at Maunakea, are responsible for some of these candidates. The right panel of Figure \ref{fig:850_FAP} plots the sinusoidal vs linear FAPs for all the monitored JCMT Transient sources over the 6 year window using the updated calibration, revealing a cluster of sources with FAPs $\sim 10^{-4}$.

Considering only the newly calibrated results, we find that 12 of the 13 robust secular variables after 4 years remain robust after 6 years of observations, with only Serpens Main SMM\,10 missing the cut (Table \ref{tab:FAP}). The FAP for SMM\,10 increases from $10^{-5.1}$ to  $10^{-4.3}$, dropping it to candidate status. Interestingly, SMM\,10 was found to have significant stochasticity in its lightcurve when observed with higher angular resolution using the ALMA ACA \citep{Francis22}. Furthermore, of the 9 candidate secular variables found after 4 years, 4 are raised to robust after 6 years, while 3 remain candidates and 2 are removed from the variable sample.

\begin{figure}
\plotone{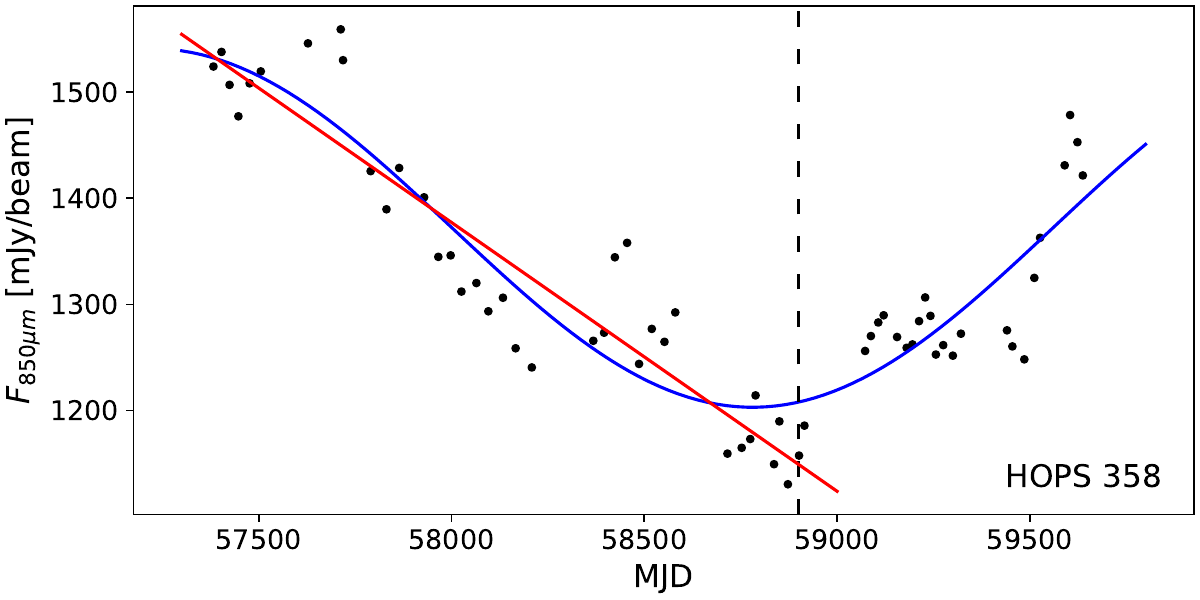}
\caption{Lightcurve of HOPS 358, which changed its variable type from linear (red) to curved (blue) after adding two additional years of observation. The dashed line marks the date of the last measurement used by \citet{yhlee2021} in their linear fit.}
\label{fig:4-6-yr-curve}
\end{figure}

The majority of the robust sources maintain the same periodic/curved/linear classification over both time windows, although the additional epochs do occasionally change the best-fit period. As an example, Figure \ref{fig:4-6-yr-curve} plots the old (red) and new (blue) calibrated measurements and best fits for a source which changed from linear to curved-type through the addition of two more years of observation.

As noted above, one robust secular variable is not identified with a previously known protostar, J182947.6+011553 in Serpens Main. Considering near-IR variability, \citet{Kaas99} catalogued a candidate YSO, source 6 in their Table 5, about 9\arcsec\ away; however, no object is visible as a localized source in the WISE mid-IR images at this position. Additionally, the source is not in the Spitzer 24 or 4.5 micron data, nor the eHOPS catalog in IRSA, suggesting that it may be a variable background galaxy. Looking further afield, SMM\,1 \citep{Casali93, enoch2009}, a robust secular \LongS variable source (Table \ref{tab:6yr_850}), lies about 45\arcsec\ to the south-east of J182947.6+011553. SMM\,1 is found to be a secular variable in the mid-IR \citep{contreras2020, park2021}, which has launched powerful jet and outflow \citep{Hull16} in the direction of J182947.6+011553. Consistent with the scattering surface seen in WISE images, theoretical modelling of the outflow by \citet{Liang20} suggest a $\sim 2\arcmin$ lobe length, which would entirely surround J182947.6+011553. Finally, both SMM\,1 and J182947.6+011553 show rising light curves, with $\sim$ 1.5--2\% increase per year (Table \ref{tab:6yr_850}). Thus, we hypothesize that this source, despite the significant distance, is being influenced by SMM\,1.

\subsection{Recovered \texorpdfstring{450\,$\mu$m}{450 micron} Variables at 6 yrs} 
\label{sec:results450_850}

\begin{figure}
\plotone{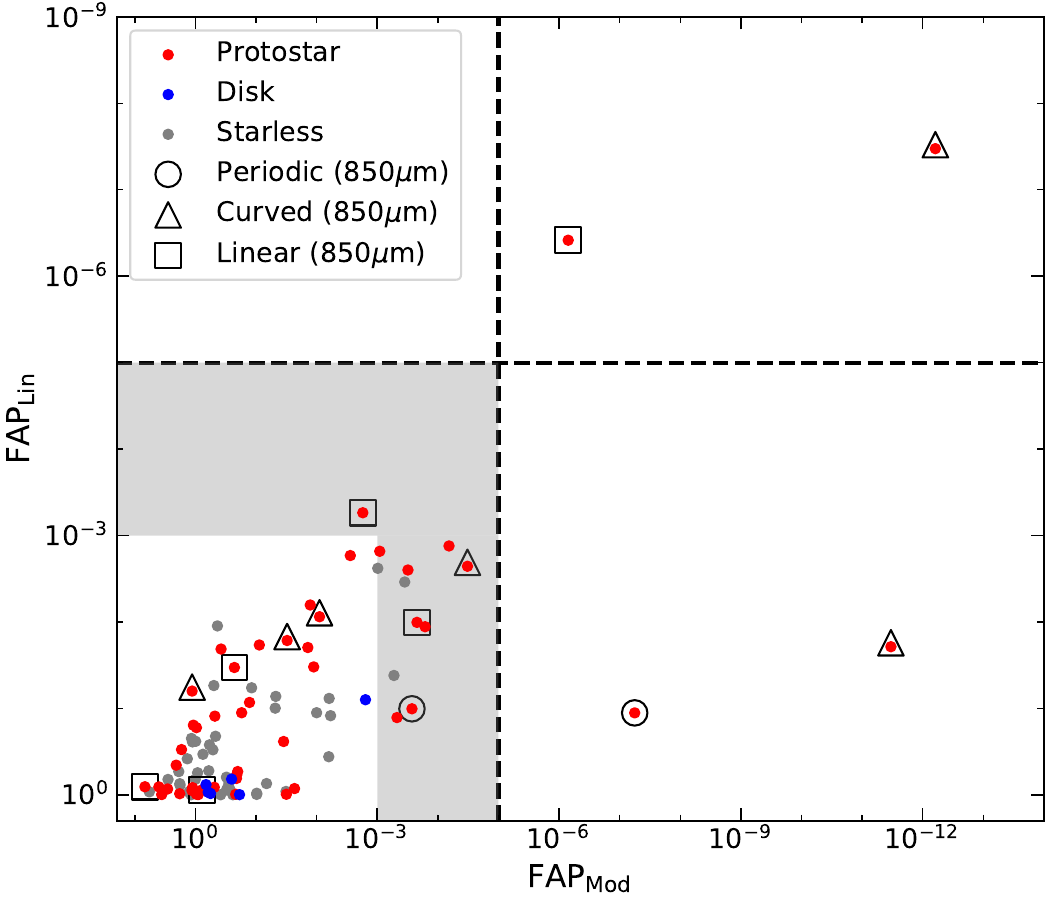}
\caption{Scatter plot of sinusoidal (FAP$_{\mathrm{Mod}}$) and linear (FAP$_{\mathrm{Lin}}$) at 450 microns. Sources that are robust variables at 850 microns are marked as circles (Periodic), triangles (Curved), and squares (Linear).}
\label{fig:450_FAP}
\end{figure}

\begin{figure*}
\plotone{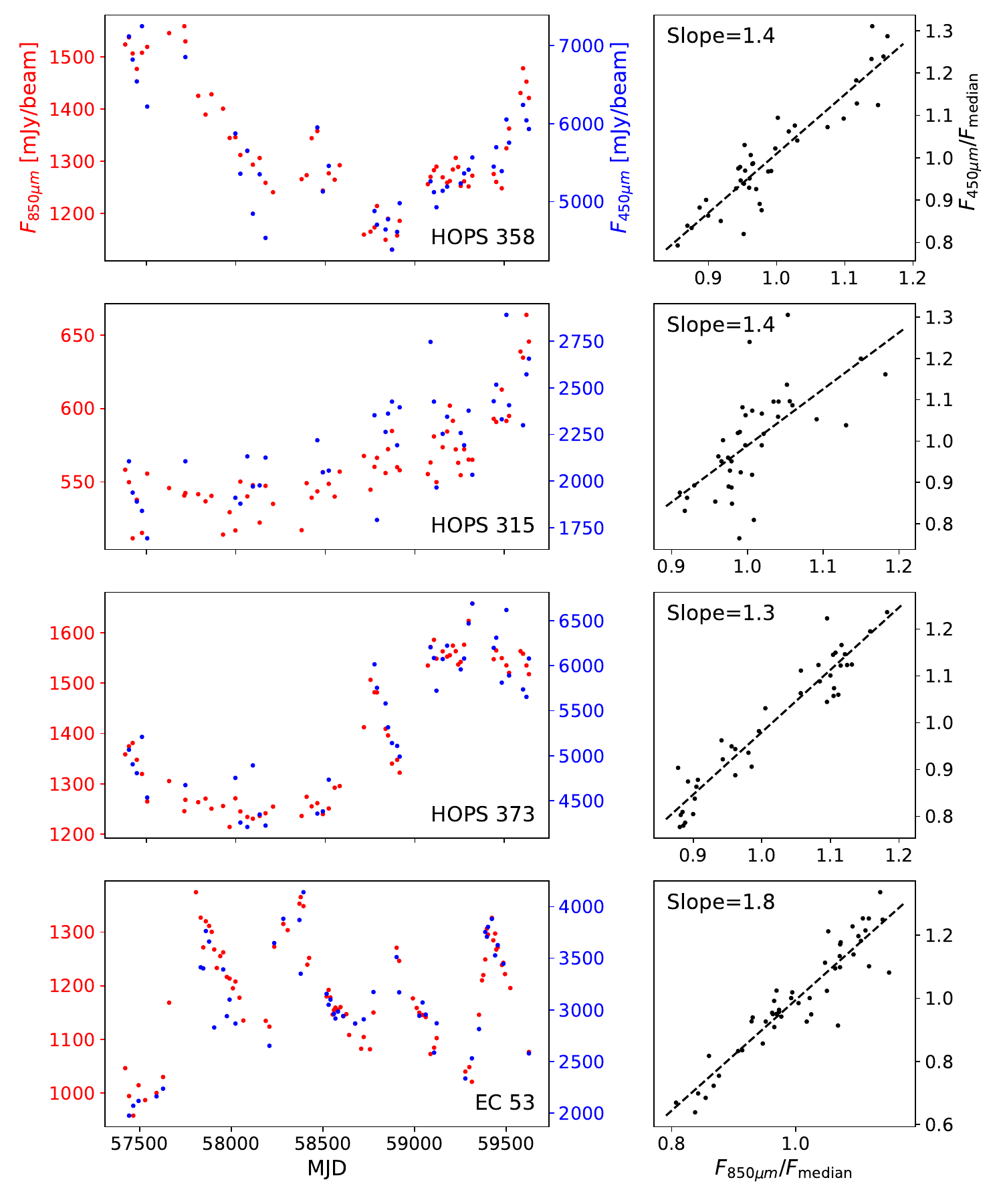}
\caption{Left panels: Light curves at 850\,$\mu$m (red dots) and 450\,$\mu$m (blue dots) for the four robust variables at both wavelengths, showing the tight correlation. Right Panels: Scatter plots of the brightness at each wavelength normalized at 850\,$\mu$m by the median measured value and at 450\,$\mu$m by the intercept value of the best-fit slope at the median 850\,$\mu$m value. The normalized slopes are provided for each source.}
\label{fig:450_850_LC}
\end{figure*}

Figure \ref{fig:450_FAP} plots the \ShortS sinusoidal vs linear FAPs for all the monitored JCMT Transient sources. There are 4 robust secular detections, all of which are also robust at \LongS (see Tables \ref{tab:FAP} and \ref{tab:6yr_450}). Furthermore, of the 12 additional candidate variables at \ShortNS, 4 are known robust variables at \LongNS. In Figure \ref{fig:450_850_LC} we show the \ShortS and \LongS light curves for the four robust at \ShortS protostellar sources, along with a scatter plot of the paired submillimetre fluxes across all epochs, revealing the linearity in the response between wavelengths. This is expected if the underlying process is a change in the temperature of the dust in the envelope due to variations in the accretion luminosity of the deeply embedded protostar \citep{johnstone2013, contreras2020,Francis22}. 

Quantifying the response at \LongS versus \ShortNS, we have calculated the normalized slopes for each of these four sources (see Figure \ref{fig:450_850_LC}, right panels). For the three Orion HOPS secular variables, the normalized slope is $\sim 1.4$, such that the \ShortS brightness has about a 40\% larger variation than the \LongS brightness. Following the same argument as used by \citet[][their Section 6.2]{contreras2020}, we anticipate that the \LongS brightness varies as $T_d^{\sim 1.5}$ while the \ShortS brightness varies as $T_d^{\sim 2}$, where we assume $T_d \sim 20\,$K in the outer envelope. The larger exponent at \ShortS is due to the fact that the shorter wavelength is less fully on the Rayleigh-Jeans tail of the dust emission and therefore has a stronger reaction to temperature change. These two formulae can be combined to yield $S_{450} \propto S_{850}^{4/3}$, which for small variations can be made linear such that $S_{450} \sim 1.33\,S_{850}$. 

The Serpens Main secular variable protostar EC\,53 \citep{yhlee2020,Francis22}, also known as V371\,Ser, shows a significantly higher normalized slope, $\sim 1.8$. This may indicate additional extended and non-varying emission at \LongNS, which is biasing the slope higher, or that the mean dust temperature $T_d$ used in the above determinations of submillimetre brightness response has been overestimated. The power-law exponents rise at both \LongS and \ShortS as the dust temperature decreases, but more strongly at \ShortNS, leading to a greater response at \ShortS versus \LongNS. We note, however, that modeling of the EC\,53 envelope structure and temperature profile by \citet{Francis22} to fit simultaneously time-variable observations at interferometric and single-dish angular scales required a somewhat higher outer dust temperature, $T_d \sim25\,$K. \citet{Francis22} also struggled to fit well these \ShortS JCMT observations (see their Section 6.2) and suggest that one complication may be the JCMT beam structure at \ShortNS. The good fits, however, to the simple \citet{contreras2020} model for the three HOPS sources suggest that the problem may lie with the simplified modeled structure assumed within the EC\,53 envelope. Thus, time-dependent calculations using a more detailed and careful radiative transfer modeling of the envelope, such as those considered by \citet{Baek20} and including the known outflow cavity and disk, are likely to be required.

\begin{figure*}
\plotone{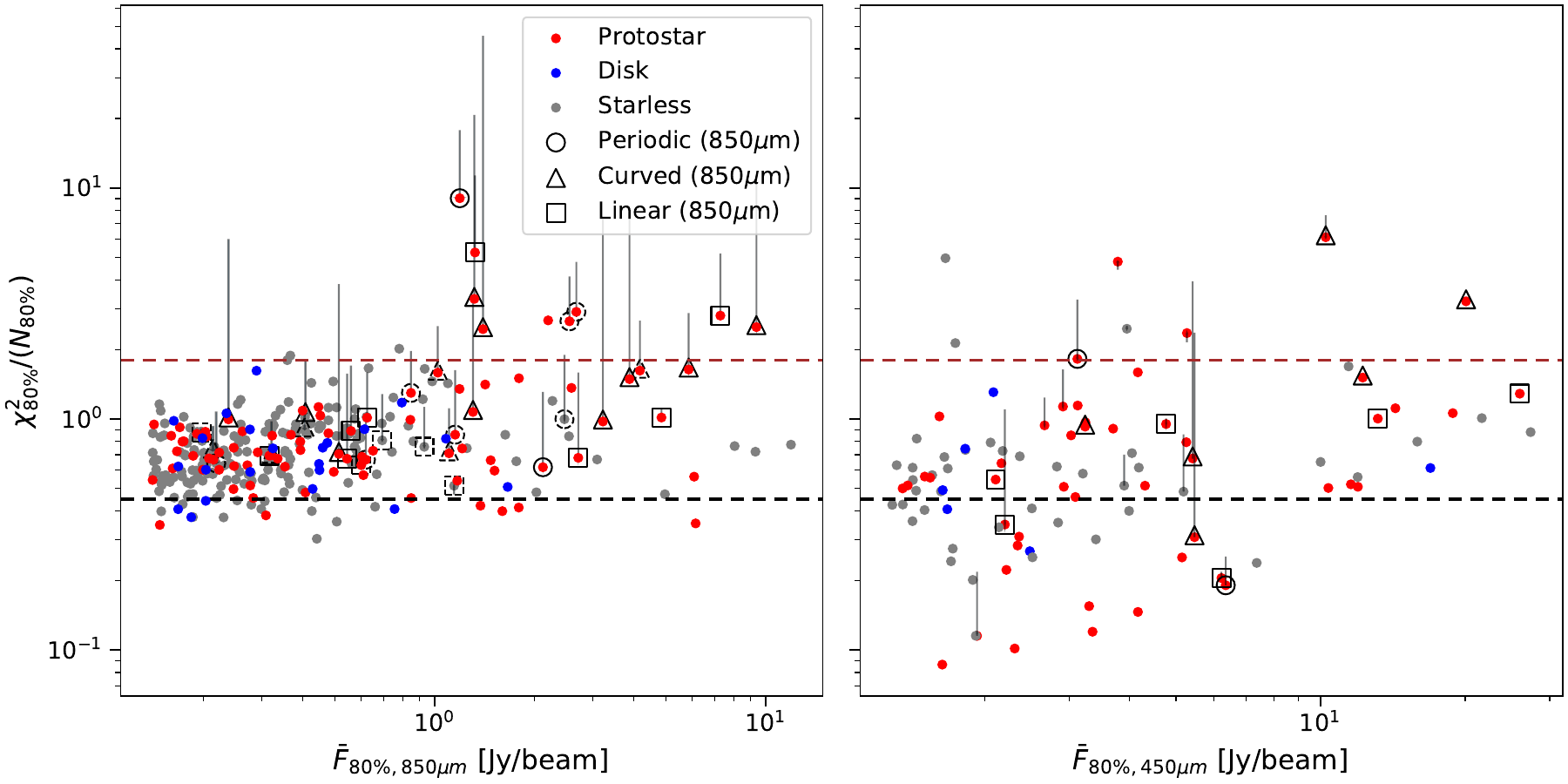}
\caption{Scatter plot of chi-square-fit value divided by the number of observations for each source. The left figure is drawn at 850 microns, and the right figure is drawn at 450 microns. Here we only include the middle 80\% data points in brightness and use Equation \ref{eq:SDfid} for the standard deviation of each measurement, with RFCF$_{\rm unc}$ equal to 1\% and 5\% at 850 and 450 microns, respectively.  The robust variables at 850 microns are also annotated. The gray vertical lines shows the change of chi-square value after the subtraction of the best-fit secular trend.}
\label{fig:chi2}
\end{figure*}

Finally, as a check on the utility of the robust secular fits at both \LongS and \ShortNS, in Figure \ref{fig:chi2} we present scatter plots at \LongS and \ShortS of the reduced $\chi^2$ versus mean flux (using the middle 80\% of data points in brightness). The measurements are made both before and after removal of the best-fit secular component, for those sources which are robust secular variables at \LongS over 6 years. As found by \citet{yhlee2021}, the robust variables tend to have exceptionally high $\chi^2$ values prior to the removal of the best secular fit. For the \LongS data sets, even the removal of the secular fit can leave a significant residual, suggesting that for some sources there exists an additional component beyond the simple smooth evolution in time of the light curves investigated here.

\begin{deluxetable*}{llccccc}
\tablecaption{Physical Properties of Robust Secular Variables at 450 Microns over 6 Years\label{tab:6yr_450}}
\small
\tablehead{
 \colhead{ID} & 
 \colhead{Known Name} & 
 \colhead{$\mathrm{\Bar{F}}$}  & 
 \colhead{Group} &
 \colhead{Period} & 
 \colhead{A/$\mathrm{\Bar{F}}$} & 
  \colhead{Slope/$\mathrm{\Bar{F}}$} \\
& & [Jy bm$^{-1}$]& & [yr]& [$\%$ yr$^{-1}$] & [$\%$] 
}
\startdata
J054607.2-001332& NGC \,2068 HOPS\,358   &5.5 &C& 12.& 28.&      \\
J054603.6-001447& NGC \,2068 HOPS\,315   &2.2 &L& 50.&    &  4.4 \\
J054631.0-000232& NGC \,2068 HOPS\,373   &5.4 &C& 8.5& 17.&  5.3 \\
J182951.2+011638& Serpens Main EC\,53    &3.1 &P& 1.5& 20.&      
\enddata
\end{deluxetable*}

\section{Summary and Conclusions} 
\label{sec:conclude}

The JCMT Transient Survey has been monitoring eight Gould Belt regions continuously since December 2015 and six intermediate mass star-forming regions since February 2020. We have reduced and calibrated all of these observations through February 2022 using \textit{Pipeline v2}, which we describe in this paper. We have also investigated the Gould Belt fields for variable source, comparing the 4 year results using \textit{Pipeline v2} against the analysis by \citet{yhlee2021} over the same time range using \textit{Pipeline v1}. We further show the importance of increasing the  monitoring observations time window by comparing the variability results using \textit{Pipeline v2} over the full 6 years of individual epochs.  Finally, we demonstrate the value of multi-wavelength monitoring by performing \textit{Pipeline v2} relative calibration on the \ShortS data and comparing individual source light curves at both \LongS and \ShortNS. The main results of this paper are therefore:
\begin{enumerate}
    \item The \textit{Pipeline v2} relative alignment via cross correlation achieves an accuracy better than $0.5^{\prime \prime}$, which can be compared directly against the \LongS and \ShortS beam sizes of $14.1^{\prime \prime}$ and $9.6^{\prime \prime}$, respectively. While the alignment is performed at \LongNS, the results are directly applicable to the \ShortS maps as the JCMT SCUBA-2 observations are carried out simultaneously by means of a dichroic beam-splitter (Section\ \ref{sec:resultalign}).

    \item The \textit{Pipeline v2} relative flux calibration uses information on all sources in each field and an iterative approach to more robustly bring all epochs of a given region into agreement. For most fields the uncertainty in the \LongS Relative-FCF ($\sigma_{\mathrm{FCF}}$) is 1\%  and at \ShortNS, the Relative-FCF uncertainty is $< 5$\% (Section \ref{sec:methodlocalfluxcal}). Furthermore, given that the JCMT Transient Survey observations are optimized for \LongS observations, we automate determination of the ``good" \ShortS epochs by applying limits to the quantified metadata (Section \ref{subsec:450data}).

    \item Along with this paper we make available to the community deep co-adds of the eight Gould Belt star-forming regions (Section \ref{sec:coadds}). The typical RMS for these maps at \LongS and \ShortS are 1.5\,mJy\,bm$^{-1}$ and 24\,mJy\,bm$^{-1}$, respectively (Table \ref{tab:CoaddGB}). We also present the reduction and calibration metadata for all Gould Belt epochs (Appendix \ref{app:GB-summary}) and all intermediate mass star-forming regions (Appendix \ref{app:HM-summary}). 

    \item We analyse the variable source results at \LongS using both \textit{Pipeline v2} and \textit{Pipeline v1} over 4 years, uncovering a greater number of robust variables using the updated calibration techniques (Section \ref{sec:results850_4yr}). The properties of the recovered variables found by both calibration methods agree in general, though there are subtleties when determining the best-fit periods via periodogram analysis. We also demonstrate that extending the time baseline to 6 years increases the number of variables recovered (Section \ref{sec:results850_6yr}).

    \item Finally, we analyse the variable source results at \ShortS for the first time. Over 6 years we recover 4 robust variables, all of which are also robust at \LongNS. Direct comparison of the light curves at both wavelengths supports the expectation that the variability is driven by changing mass accretion onto the central protostar and the resultant heating of the dust in the natal envelope (Section \ref{sec:results450_850}).

\end{enumerate}

Co-added images are available here: \url{https://www.canfar.net/citation/landing?doi=23.0009}.

\section{acknowledgements}

The authors wish to recognize and acknowledge the
very significant cultural role and reverence that the summit of Maunakea has always had within the indigenous
Hawaiian community. We are most fortunate to have the
opportunity to conduct observations from this mountain. 
The comments provided by the anonymous referee have significantly strengthened this work.
The James Clerk Maxwell Telescope is operated by the 
East Asian Observatory on behalf of The National 
Astronomical Observatory of Japan; Academia Sinica Institute of Astronomy and Astrophysics; the Korea Astronomy and 
Space Science Institute; the National Astronomical 
Research Institute of Thailand; Center for 
Astronomical Mega-Science (as well as the National Key 
R\&D Program of China with No. 2017YFA0402700). Additional 
funding support is provided by the Science and Technology 
Facilities Council of the United Kingdom and participating 
universities and organizations in the United Kingdom and Canada. Additional funds for the construction
of SCUBA-2 were provided by the Canada Foundation
for Innovation. This research used the facilities of the
Canadian Astronomy Data Centre operated by the National Research  Council of Canada with the support of
the Canadian Space Agency. The Starlink
software \citep{currie2014} is currently supported by
the East Asian Observatory.

The full JCMT Transient Team includes the authors of this paper, as well as the authors of \citet{yhlee2021} and \citet{herczeg2017}. D.J.\ is supported by NRC Canada and by an NSERC Discovery Grant. G.P. is supported by the National Research Foundation of Korea through grants NRF-2020R1A6A3A01100208 \& RS-2023-00242652.


\facilities{JCMT}

\software{astropy \citep{astropy},  
          matplotlib \citep{matplotlib},
          aplpy \citep{aplpy}
          starlink \citep{currie2014}, 
          }
\newpage

\bibliographystyle{aasjournal}
\bibliography{CalPipelinev2}


\appendix

\section{Summary of Gould Belt Observations}
\label{app:GB-summary}

Table \ref{tab:GB-obs-summary} summarizes the Gould Belt observations 
taken since the beginning of the JCMT Transient Survey (2015 December) 
through 2022 February. The pointing offset corrections are given by 
$\Delta\mathrm{R.A.}$ and $\Delta\mathrm{Dec}$. The background RMS values
were measured before the \textit{Pipeline v2} Relative FCFs ($R_{\mathrm{FCF}}$) 
were applied. $C_{\lambda}$ is the ratio of the \textit{Pipeline v1} and 
\textit{Pipeline v2} $R_{\mathrm{FCF}}$, as shown in the right panels of Figures 
\ref{fig:FluxCalCompare850} and \ref{fig:FluxCalCompare450}. The 
observations that have an associated $R_{\mathrm{FCF},450}$ are the 
``usable'' 450\,$\mu$m maps, as defined in Section \ref{subsec:450data}. 
In the case of IC\,348, an $R_{\mathrm{FCF},450}$ value is often given, 
while the $C_{450}$ ratio could not be calculated. This indicates that a 
$R_{\mathrm{FCF},450}$ factor could not be derived using \textit{Pipeline v1} 
because the ``Point-Source method'' algorithm relies on the the flux 
measurements of only two sources and one source is near enough the 
minimum brightness threshold that a robust peak flux measurement cannot 
be performed. 

Public JCMT Transient Survey Data can be found at the Canadian Astronomy Data Centre 
(\textit{CADC})\footnote{\url{https://www.cadc-ccda.hia-iha.nrc-cnrc.gc.ca/}} by searching for 
proposal IDs M16AL001 and M20AL007. Note that region NGC\,2068 is listed in the archive under the name
NGC\,2071.

\begin{center}
{\small\tabcolsep=3pt  
\begin{longtable}{ >{\bfseries\footnotesize}cccccccccccc}
\caption{Summary of JCMT Transient Survey Gould Belt Regions Processed Observations}
\label{tab:GB-obs-summary} \\ 
\hline
\multicolumn{1}{|c|}{\textbf{Region}} &
\multicolumn{1}{c|}{\textbf{Date}} &
\multicolumn{1}{c|}{\textbf{Scan}} &
\multicolumn{1}{c|}{\textbf{$\tau\times\mathrm{AM}$}} &
\multicolumn{1}{c|}{\textbf{RMS$_{450}$}} &
\multicolumn{1}{c|}{\textbf{RMS$_{850}$}} &
\multicolumn{1}{c|}{\textbf{$\Delta$R.A.}} &
\multicolumn{1}{c|}{\textbf{$\Delta$Dec}} &
\multicolumn{1}{c|}{\textbf{$R\mathrm{-FCF}_{\mathrm{450}}$}} &
\multicolumn{1}{c|}{\textbf{$R\mathrm{-FCF}_{\mathrm{850}}$}} &
\multicolumn{1}{c|}{\textbf{C$_{\mathrm{450}}$}} &
\multicolumn{1}{c|}{\textbf{C$_{\mathrm{850}}$}}\\ 
\multicolumn{1}{|c|}{} &
\multicolumn{1}{c|}{} &
\multicolumn{1}{c|}{} &
\multicolumn{1}{c|}{} &
\multicolumn{1}{c|}{\textbf{(mJy/beam)}} &
\multicolumn{1}{c|}{\textbf{(mJy/beam)}} &
\multicolumn{1}{c|}{\textbf{(arcsec)}} &
\multicolumn{1}{c|}{\textbf{(arcsec)}} &
\multicolumn{1}{c|}{} &
\multicolumn{1}{c|}{} &
\multicolumn{1}{c|}{} &
\multicolumn{1}{c|}{}
\\ \hline
\endfirsthead
\multicolumn{5}{c}%
{{\bfseries \tablename\ \thetable{} -- continued from previous page}} \\ 
\hline
 \multicolumn{1}{|c|}{\textbf{Region}} &
\multicolumn{1}{c|}{\textbf{Date}} &
\multicolumn{1}{c|}{\textbf{Scan}} &
\multicolumn{1}{c|}{\textbf{$\tau\times\mathrm{AM}$}} &
\multicolumn{1}{c|}{\textbf{RMS$_{450}$}} &
\multicolumn{1}{c|}{\textbf{RMS$_{850}$}} &
\multicolumn{1}{c|}{\textbf{$\Delta$R.A.}} &
\multicolumn{1}{c|}{\textbf{$\Delta$Dec}} &
\multicolumn{1}{c|}{\textbf{$R_{\mathrm{FCF,450}}$}} &
\multicolumn{1}{c|}{\textbf{$R_{\mathrm{FCF,850}}$}} &
\multicolumn{1}{c|}{\textbf{C$_{\mathrm{450}}$}} &
\multicolumn{1}{c|}{\textbf{C$_{\mathrm{850}}$}}\\ 
\multicolumn{1}{|c|}{} &
\multicolumn{1}{c|}{} &
\multicolumn{1}{c|}{} &
\multicolumn{1}{c|}{} &
\multicolumn{1}{c|}{\textbf{(mJy/beam)}} &
\multicolumn{1}{c|}{\textbf{(mJy/beam)}} &
\multicolumn{1}{c|}{\textbf{(arcsec)}} &
\multicolumn{1}{c|}{\textbf{(arcsec)}} &
\multicolumn{1}{c|}{} &
\multicolumn{1}{c|}{} &
\multicolumn{1}{c|}{} &
\multicolumn{1}{c|}{}
\\ \hline
\endhead
\hline
\multicolumn{3}{|r|}{{Continued on next page}} \\ 
\hline
\endfoot
\hline \hline
\endlastfoot
IC 348 & 2015-12-22 & 19 & 0.08 & 161.74 & 11.62 & 2.35 & 1.16 & 1.0 $\pm$ 0.025 & 0.92 $\pm$ 0.021 & -- & 0.979 \\ 
IC 348 & 2016-01-15 & 22 & 0.08 & 135.76 & 9.37 & 4.74 & 2.16 & 0.92 $\pm$ 0.023 & 0.93 $\pm$ 0.017 & -- & 0.981 \\ 
IC 348 & 2016-02-05 & 18 & 0.04 & 72.81 & 12.07 & 4.47 & 5.51 & 0.71 $\pm$ 0.018 & 0.9 $\pm$ 0.022 & -- & 0.985 \\ 
IC 348 & 2016-02-26 & 20 & 0.08 & 146.27 & 11.3 & 4.45 & 4.61 & 0.8 $\pm$ 0.034 & 0.9 $\pm$ 0.016 & -- & 0.985 \\ 
IC 348 & 2016-03-18 & 27 & 0.08 & 126.86 & 10.2 & 7.46 & 2.26 & 0.79 $\pm$ 0.031 & 0.87 $\pm$ 0.018 & -- & 0.983 \\ 
IC 348 & 2016-04-17 & 9 & 0.07 & 85.18 & 10.34 & 5.17 & 5.75 & 0.7 $\pm$ 0.017 & 0.84 $\pm$ 0.014 & -- & 1.013 \\ 
IC 348 & 2016-08-26 & 40 & 0.13 & 473.24 & 14.03 & 6.16 & 0.5 & -- & 0.88 $\pm$ 0.017 & -- & 0.99 \\ 
IC 348 & 2016-11-26 & 22 & 0.08 & 128.83 & 11.5 & 5.92 & 2.29 & 1.29 $\pm$ 0.032 & 1.09 $\pm$ 0.011 & -- & 1.004 \\
\end{longtable}}
\end{center}

\section{Summary of Intermediate-mass Regions}
\label{app:HM-summary}

\begin{figure*}
\plotone{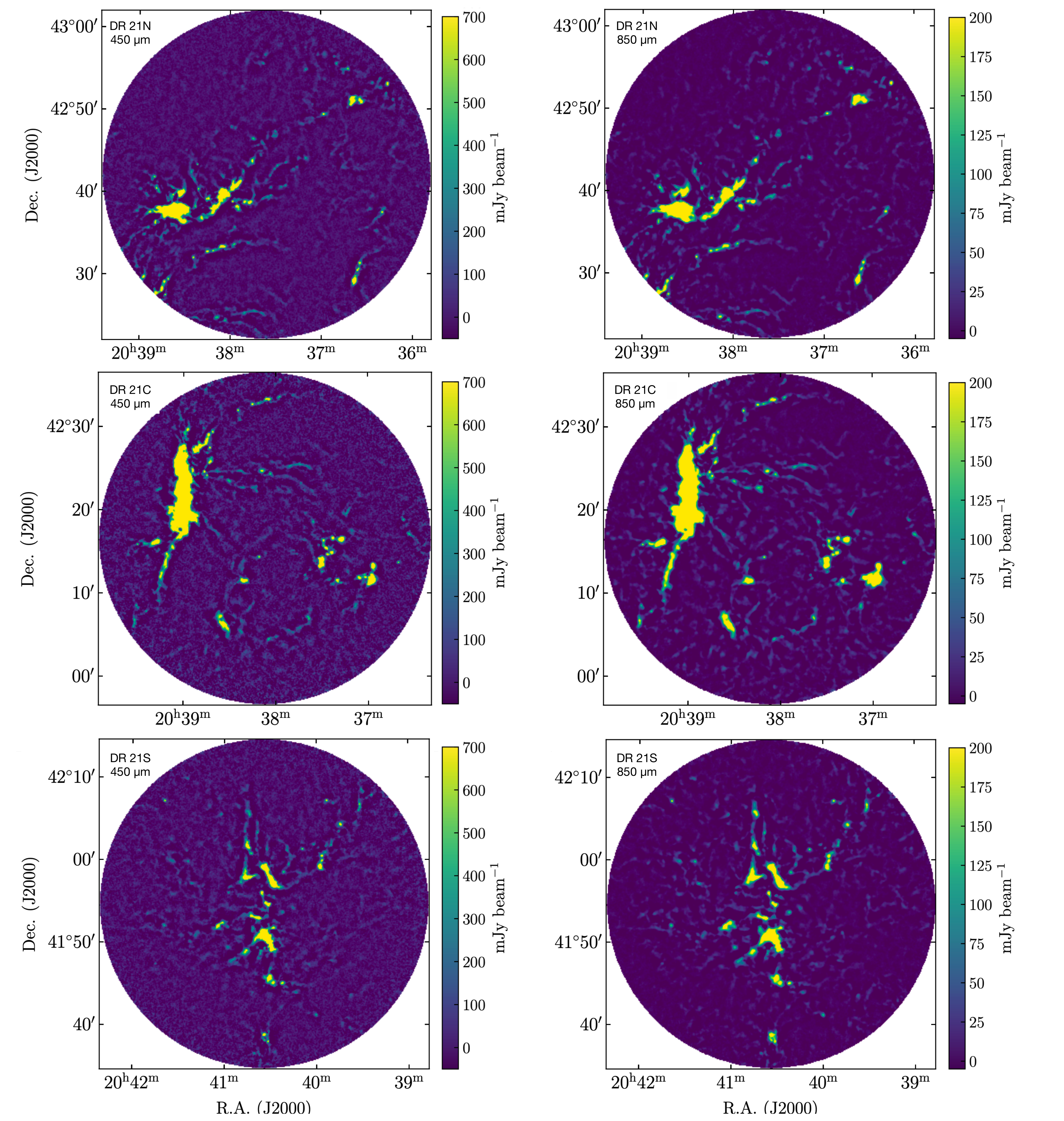}
\caption{Co-added images of DR\,21N, DR\,21C, and DR\,21S (top to bottom, respectively) at both 450 (left) and 850$\,\mu$m (right). Observations through February 2022 are included. At 450$\,\mu$m, only ``usable'' data as defined in Section \ref{subsec:450data} are included.}
\label{fig:CoaddsDR21}
\end{figure*}

\begin{figure*}
\plotone{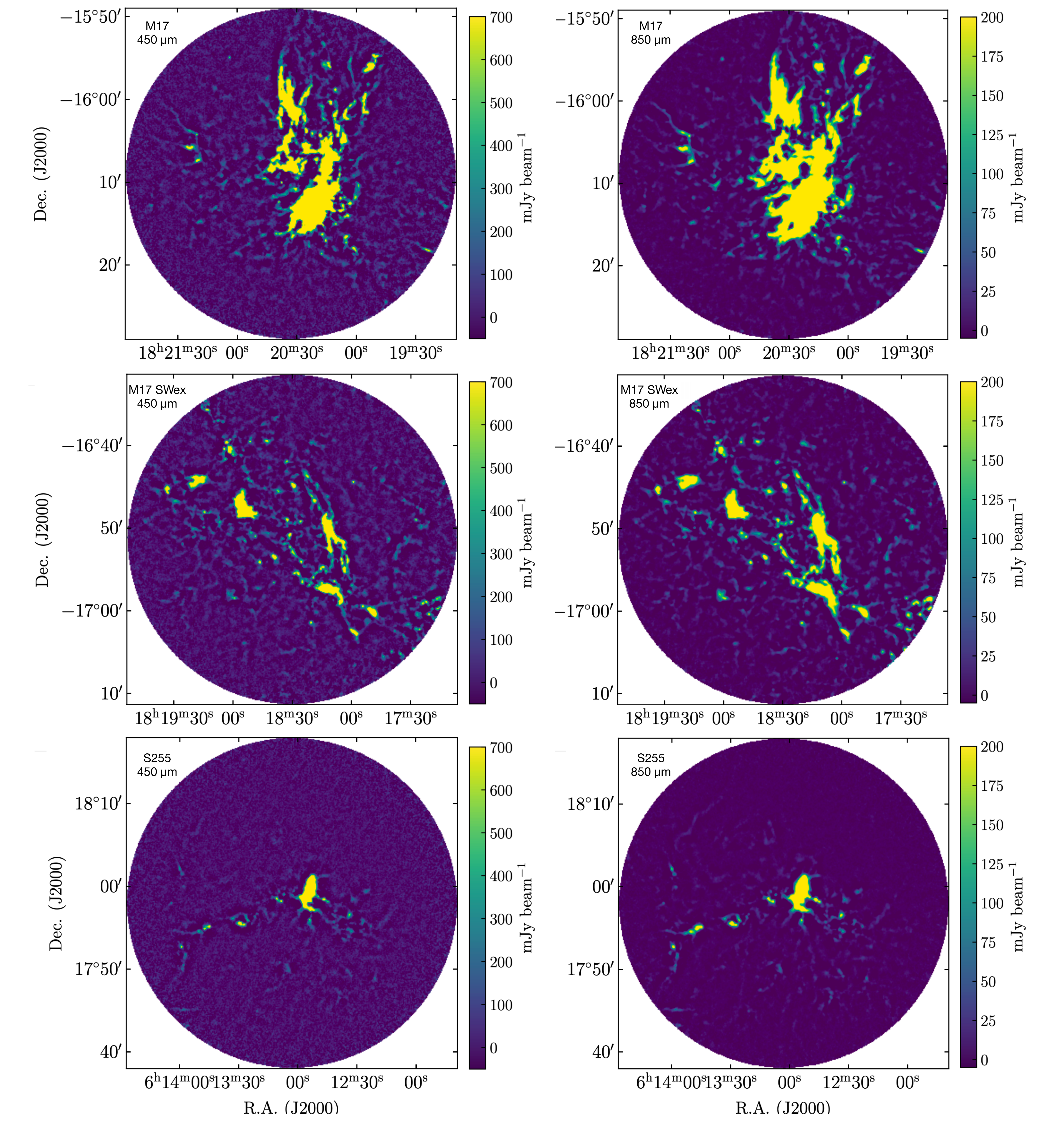}
\caption{Co-added images of M17, M17\,SWex, and S255 (top to bottom, respectively) at both 450 (left) and 850$\,\mu$m (right). Observations through February 2022 are included. At 450$\,\mu$m, only ``usable'' data as defined in Section \ref{subsec:450data} are included.}
\label{fig:CoaddsM17S255}
\end{figure*}

In February 2020, the JCMT Transient survey was expanded to 
monitor six fields outside the Gould Belt and at higher 
distances, toward regions of intermediate/high mass 
star-formation: three fields in DR21 (North, Central, 
South), M17, M17\,SWex, and S255. Co-added images of each
region are presented in Figures \ref{fig:CoaddsDR21} and
\ref{fig:CoaddsM17S255}. Tables \ref{tab:CoaddHM} and 
\ref{tab:HM-obs-summary} summarize the properties of the 
co-added and individual images obtained, respectively. 

As noted in the main text, these fields have 
significantly fewer observations than the original eight 
Gould Belt targets, requiring a separate, more detailed 
analysis of the calibration uncertainty and source 
variability. Further information tailored to each new region 
along with first variability results in each region which 
will be released in a separate publication.

\clearpage

\begin{deluxetable}{ccccccccc}
\tablecaption{Summary of Intermediate-mass region co-add images.}
\label{tab:CoaddHM}
\tablecolumns{9}
\tablewidth{0pt}
\tablehead{
\colhead{Region} &
\colhead{R.A.} &
\colhead{Dec} &
\colhead{RMS$_{450}$} &
\colhead{RMS$_{850}$} &
\colhead{N$_{450}$}  &
\colhead{N$_{850}$} &
\colhead{C$_{450}$} &
\colhead{C$_{850}$} \\
\colhead{} &
\colhead{(J2000)} &
\colhead{(J2000)} &
\colhead{(mJy\,bm$^{-1}$)} &
\colhead{(mJy\,bm$^{-1}$)} &
\colhead{}  &
\colhead{} &
\colhead{(SD)} &
\colhead{(SD)}
}
\startdata
DR21-C   & 20:38:09 & +42:16:39 & 44 & 2.4 & 12 & 18 & X & X \\
DR21-N   & 20:37:36 & +42:41:42 & 38 & 2.5 & 10 & 17 & X & X \\
DR21-S   & 20:40:35 & +41:54:54 & 35 & 2.4 & 10 & 18 & X & X \\
M17      & 18:20:31 & -16:08:54 & 29 & 2.6 &  9 & 14 & X & X \\
M17SWex  & 18:18:28 & -16:51:35 & 20 & 2.8 &  8 & 13 & X & X \\
S255     & 06:13:04 & +17:58:32 & 36 & 2.6 &  9 & 14 & X & X \\
\enddata
\end{deluxetable}

\begin{center}
{\small\tabcolsep=3pt  
\begin{longtable}{ >{\bfseries\footnotesize}cccccccccccc}
\caption{Summary of JCMT Transient Survey Intermediate-Mass Region Processed Observations}
\label{tab:HM-obs-summary} \\ 
\hline
\multicolumn{1}{|c|}{\textbf{Region}} &
\multicolumn{1}{c|}{\textbf{Date}} &
\multicolumn{1}{c|}{\textbf{Scan}} &
\multicolumn{1}{c|}{\textbf{$\tau\times\mathrm{AM}$}} &
\multicolumn{1}{c|}{\textbf{RMS$_{450}$}} &
\multicolumn{1}{c|}{\textbf{RMS$_{850}$}} &
\multicolumn{1}{c|}{\textbf{$\Delta$R.A.}} &
\multicolumn{1}{c|}{\textbf{$\Delta$Dec}} &
\multicolumn{1}{c|}{\textbf{$R\mathrm{-FCF}_{\mathrm{450}}$}} &
\multicolumn{1}{c|}{\textbf{$R\mathrm{-FCF}_{\mathrm{850}}$}} &
\multicolumn{1}{c|}{\textbf{C$_{\mathrm{450}}$}} &
\multicolumn{1}{c|}{\textbf{C$_{\mathrm{850}}$}}\\ 
\multicolumn{1}{|c|}{} &
\multicolumn{1}{c|}{} &
\multicolumn{1}{c|}{} &
\multicolumn{1}{c|}{} &
\multicolumn{1}{c|}{\textbf{(mJy/beam)}} &
\multicolumn{1}{c|}{\textbf{(mJy/beam)}} &
\multicolumn{1}{c|}{\textbf{(arcsec)}} &
\multicolumn{1}{c|}{\textbf{(arcsec)}} &
\multicolumn{1}{c|}{} &
\multicolumn{1}{c|}{} &
\multicolumn{1}{c|}{} &
\multicolumn{1}{c|}{}
\\ \hline
\endfirsthead
\multicolumn{5}{c}%
{{\bfseries \tablename\ \thetable{} -- continued from previous page}} \\ 
\hline
 \multicolumn{1}{|c|}{\textbf{Region}} &
\multicolumn{1}{c|}{\textbf{Date}} &
\multicolumn{1}{c|}{\textbf{Scan}} &
\multicolumn{1}{c|}{\textbf{$\tau\times\mathrm{AM}$}} &
\multicolumn{1}{c|}{\textbf{RMS$_{450}$}} &
\multicolumn{1}{c|}{\textbf{RMS$_{850}$}} &
\multicolumn{1}{c|}{\textbf{$\Delta$R.A.}} &
\multicolumn{1}{c|}{\textbf{$\Delta$Dec}} &
\multicolumn{1}{c|}{\textbf{$R_{\mathrm{FCF,450}}$}} &
\multicolumn{1}{c|}{\textbf{$R_{\mathrm{FCF,850}}$}} &
\multicolumn{1}{c|}{\textbf{C$_{\mathrm{450}}$}} &
\multicolumn{1}{c|}{\textbf{C$_{\mathrm{850}}$}}\\ 
\multicolumn{1}{|c|}{} &
\multicolumn{1}{c|}{} &
\multicolumn{1}{c|}{} &
\multicolumn{1}{c|}{} &
\multicolumn{1}{c|}{\textbf{(mJy/beam)}} &
\multicolumn{1}{c|}{\textbf{(mJy/beam)}} &
\multicolumn{1}{c|}{\textbf{(arcsec)}} &
\multicolumn{1}{c|}{\textbf{(arcsec)}} &
\multicolumn{1}{c|}{} &
\multicolumn{1}{c|}{} &
\multicolumn{1}{c|}{} &
\multicolumn{1}{c|}{}
\\ \hline
\endhead
\hline
\multicolumn{3}{|r|}{{Continued on next page}} \\ 
\hline
\endfoot
\hline \hline
\endlastfoot
DR21-C & 2020-02-22 & 88 & 0.04 & 56.94 & 9.77 & 0.67 & 0.0 & 1.0 $\pm$ 0.01 & 1.06 $\pm$ 0.113 & 1.029 & 1.001 \\ 
DR21-C & 2020-05-21 & 56 & 0.14 & 506.17 & 11.32 & 4.72 & -1.42 & -- & 0.97 $\pm$ 0.004 & -- & 0.987 \\ 
DR21-C & 2020-06-21 & 20 & 0.12 & 286.79 & 9.63 & 3.68 & 2.67 & -- & 1.12 $\pm$ 0.003 & -- & 1.003 \\ 
DR21-C & 2020-07-30 & 44 & 0.05 & 80.95 & 12.54 & 2.77 & -1.42 & 1.0 $\pm$ 0.006 & 1.03 $\pm$ 0.004 & 1.01 & 0.999 \\ 
DR21-C & 2020-09-01 & 25 & 0.05 & 48.28 & 7.52 & 1.71 & 1.96 & 0.96 $\pm$ 0.005 & 1.01 $\pm$ 0.003 & 1.021 & 1.0 \\ 
DR21-C & 2020-10-08 & 23 & 0.12 & 357.81 & 11.21 & 2.53 & 1.35 & 1.05 $\pm$ 0.032 & 0.98 $\pm$ 0.005 & 0.94 & 0.989 \\ 
DR21-C & 2020-11-27 & 19 & 0.13 & 388.5 & 10.79 & 0.32 & -2.02 & -- & 0.95 $\pm$ 0.004 & -- & 1.005 \\ 
DR21-C & 2021-02-08 & 88 & 0.09 & 158.35 & 10.28 & 1.06 & 0.71 & 1.76 $\pm$ 0.009 & 1.12 $\pm$ 0.003 & 1.021 & 1.004 \\ 
DR21-C & 2021-03-26 & 74 & 0.08 & 109.86 & 9.42 & 1.01 & -0.2 & 0.98 $\pm$ 0.008 & 0.99 $\pm$ 0.003 & 1.031 & 1.01 \\ 
DR21-C & 2021-03-26 & 75 & 0.07 & 90.74 & 8.9 & 0.95 & 0.23 & 0.94 $\pm$ 0.006 & 1.0 $\pm$ 0.003 & 1.032 & 1.01 \\ 
DR21-C & 2021-04-23 & 57 & 0.14 & 546.44 & 11.4 & 2.31 & -0.93 & -- & 0.97 $\pm$ 0.004 & -- & 0.998 \\ 
DR21-C & 2021-05-29 & 44 & 0.12 & 310.1 & 10.5 & 1.34 & -1.27 & 1.11 $\pm$ 0.024 & 0.97 $\pm$ 0.004 & 0.954 & 1.0 \\ 
DR21-C & 2021-06-29 & 24 & 0.09 & 124.97 & 10.61 & 0.65 & 3.49 & 1.0 $\pm$ 0.01 & 1.0 $\pm$ 0.004 & 0.992 & 0.998 \\ 
DR21-C & 2021-08-04 & 22 & 0.16 & 683.93 & 14.26 & 3.47 & 0.53 & -- & 0.93 $\pm$ 0.005 & -- & 0.997 \\ 
DR21-C & 2021-09-01 & 28 & 0.12 & 289.76 & 10.23 & 2.67 & 0.35 & 0.99 $\pm$ 0.024 & 0.97 $\pm$ 0.004 & 0.969 & 1.001 \\ 
DR21-C & 2021-09-29 & 17 & 0.07 & 119.26 & 10.21 & 0.03 & 0.54 & 0.84 $\pm$ 0.009 & 0.98 $\pm$ 0.004 & 1.021 & 0.999 \\ 
DR21-C & 2021-10-28 & 14 & 0.11 & 237.89 & 10.39 & 0.31 & 0.42 & 0.96 $\pm$ 0.018 & 0.95 $\pm$ 0.004 & 0.957 & 0.997 \\ 
DR21-C & 2021-11-26 & 18 & 0.13 & 333.71 & 10.83 & 0.56 & 2.04 & -- & 0.95 $\pm$ 0.004 & -- & 1.006 \\ 
DR21-N & 2020-02-22 & 84 & 0.04 & 54.29 & 9.47 & 2.13 & 0.0 & 0.98 $\pm$ 0.01 & 1.05 $\pm$ 0.005 & 1.041 & 1.013 \\ 
DR21-N & 2020-05-21 & 47 & 0.13 & 422.67 & 11.37 & 3.36 & -0.15 & -- & 0.96 $\pm$ 0.005 & -- & 1.004 \\ 
DR21-N & 2020-06-23 & 24 & 0.15 & 621.56 & 11.97 & 6.27 & -2.99 & -- & 0.91 $\pm$ 0.007 & -- & 0.993 \\ 
DR21-N & 2020-07-30 & 46 & 0.05 & 78.79 & 12.3 & 8.3 & -2.41 & 1.0 $\pm$ 0.006 & 1.03 $\pm$ 0.005 & 1.006 & 1.01 \\ 
DR21-N & 2020-09-02 & 27 & 0.07 & 106.54 & 14.88 & 2.85 & 0.29 & 0.96 $\pm$ 0.012 & 1.01 $\pm$ 0.006 & 1.038 & 0.994 \\ 
DR21-N & 2020-10-08 & 14 & 0.12 & 336.38 & 13.04 & 3.69 & 1.47 & -- & 0.95 $\pm$ 0.007 & -- & 1.0 \\ 
DR21-N & 2020-11-27 & 23 & 0.14 & 422.4 & 10.96 & 4.26 & 0.35 & -- & 0.94 $\pm$ 0.006 & -- & 1.017 \\ 
DR21-N & 2021-01-30 & 80 & 0.07 & 78.07 & 8.49 & 5.03 & -2.23 & 1.13 $\pm$ 0.007 & 1.03 $\pm$ 0.004 & 1.014 & 0.996 \\ 
DR21-N & 2021-03-03 & 78 & 0.05 & 47.94 & 6.94 & 4.35 & 0.07 & 1.12 $\pm$ 0.007 & 1.03 $\pm$ 0.004 & 1.039 & 1.012 \\ 
DR21-N & 2021-04-06 & 53 & 0.07 & 103.9 & 9.08 & 4.64 & 0.61 & 1.0 $\pm$ 0.01 & 1.0 $\pm$ 0.004 & 0.99 & 1.004 \\ 
DR21-N & 2021-05-17 & 81 & 0.05 & 49.2 & 7.9 & 4.49 & -0.36 & 0.9 $\pm$ 0.008 & 1.0 $\pm$ 0.005 & 0.998 & 1.007 \\ 
DR21-N & 2021-06-14 & 40 & 0.07 & 80.31 & 8.06 & 3.66 & 1.9 & 0.99 $\pm$ 0.01 & 1.0 $\pm$ 0.004 & 1.014 & 0.998 \\ 
DR21-N & 2021-07-18 & 50 & 0.14 & 411.48 & 11.48 & 3.1 & -5.41 & -- & 0.94 $\pm$ 0.007 & -- & 1.007 \\ 
DR21-N & 2021-08-15 & 13 & 0.06 & 89.65 & 9.37 & 5.87 & -0.67 & 0.9 $\pm$ 0.011 & 0.98 $\pm$ 0.006 & 1.004 & 1.006 \\ 
DR21-N & 2021-09-19 & 33 & 0.15 & 586.63 & 12.2 & 4.3 & -1.3 & -- & 0.96 $\pm$ 0.007 & -- & 0.998 \\ 
DR21-N & 2021-10-17 & 15 & 0.09 & 143.3 & 9.03 & 3.66 & 1.1 & 0.91 $\pm$ 0.015 & 0.96 $\pm$ 0.005 & 0.998 & 1.012 \\ 
DR21-N & 2021-11-19 & 16 & 0.15 & 583.32 & 12.13 & 4.77 & -1.44 & -- & 0.93 $\pm$ 0.006 & -- & 1.005 \\ 
DR21-S & 2020-02-22 & 83 & 0.04 & 55.47 & 9.38 & 1.57 & -0.82 & 1.01 $\pm$ 0.014 & 1.05 $\pm$ 0.005 & 1.013 & 1.007 \\ 
DR21-S & 2020-05-21 & 51 & 0.13 & 443.33 & 11.3 & 2.61 & -1.99 & -- & 0.97 $\pm$ 0.005 & -- & 1.015 \\ 
DR21-S & 2020-06-23 & 29 & 0.13 & 413.62 & 11.08 & 4.73 & -2.99 & -- & 0.9 $\pm$ 0.005 & -- & 1.004 \\ 
DR21-S & 2020-07-30 & 41 & 0.05 & 81.3 & 13.5 & 3.96 & -3.14 & 0.98 $\pm$ 0.007 & 1.03 $\pm$ 0.007 & 1.011 & 1.001 \\ 
DR21-S & 2020-09-02 & 21 & 0.07 & 110.06 & 11.75 & 3.39 & -0.58 & 0.97 $\pm$ 0.013 & 0.99 $\pm$ 0.006 & 1.002 & 1.002 \\ 
DR21-S & 2020-10-08 & 28 & 0.13 & 386.0 & 11.14 & 3.81 & -1.64 & -- & 0.97 $\pm$ 0.006 & -- & 0.99 \\ 
DR21-S & 2020-11-27 & 15 & 0.12 & 296.68 & 10.33 & 1.39 & -2.92 & 1.1 $\pm$ 0.024 & 0.98 $\pm$ 0.005 & 0.956 & 1.005 \\ 
DR21-S & 2020-12-30 & 19 & 0.21 & 2516.63 & 16.0 & 3.92 & -4.56 & -- & 0.89 $\pm$ 0.008 & -- & 0.987 \\ 
DR21-S & 2021-01-30 & 76 & 0.07 & 72.58 & 7.45 & 3.14 & -1.12 & 1.33 $\pm$ 0.01 & 1.09 $\pm$ 0.004 & 0.998 & 1.003 \\ 
DR21-S & 2021-03-04 & 72 & 0.06 & 100.7 & 11.31 & 2.2 & 1.37 & 1.06 $\pm$ 0.008 & 1.01 $\pm$ 0.007 & 0.989 & 1.003 \\ 
DR21-S & 2021-04-06 & 55 & 0.07 & 95.38 & 9.09 & 2.53 & -0.33 & 0.99 $\pm$ 0.008 & 1.01 $\pm$ 0.004 & 0.978 & 1.0 \\ 
DR21-S & 2021-05-17 & 87 & 0.05 & 40.2 & 7.6 & 3.62 & -1.88 & 0.94 $\pm$ 0.007 & 1.02 $\pm$ 0.004 & 1.006 & 1.002 \\ 
DR21-S & 2021-06-14 & 36 & 0.08 & 92.51 & 8.29 & 2.65 & -0.04 & 1.03 $\pm$ 0.008 & 1.01 $\pm$ 0.004 & 0.986 & 1.006 \\ 
DR21-S & 2021-07-26 & 19 & 0.09 & 163.81 & 9.27 & 3.31 & -2.79 & 0.98 $\pm$ 0.015 & 0.96 $\pm$ 0.006 & 1.005 & 1.005 \\ 
DR21-S & 2021-08-27 & 20 & 0.06 & 64.63 & 8.53 & 4.03 & -2.26 & 0.96 $\pm$ 0.009 & 1.02 $\pm$ 0.005 & 1.0 & 0.996 \\ 
DR21-S & 2021-09-27 & 16 & 0.08 & 110.41 & 8.89 & 3.44 & -0.42 & 0.75 $\pm$ 0.013 & 0.92 $\pm$ 0.004 & 1.036 & 1.01 \\ 
DR21-S & 2021-10-26 & 30 & 0.13 & 366.52 & 11.26 & 5.82 & -0.24 & -- & 0.95 $\pm$ 0.006 & -- & 1.001 \\ 
DR21-S & 2021-11-26 & 13 & 0.11 & 209.35 & 10.09 & 1.41 & -3.8 & 1.0 $\pm$ 0.028 & 0.95 $\pm$ 0.006 & -- & 1.01 \\ 
M17 & 2020-02-22 & 73 & 0.05 & 74.38 & 9.53 & -0.08 & -0.12 & 1.19 $\pm$ 0.004 & 1.08 $\pm$ 0.002 & 0.99 & 1.004 \\ 
M17 & 2020-05-21 & 42 & 0.15 & 629.39 & 11.69 & -0.34 & 0.59 & -- & 0.98 $\pm$ 0.003 & -- & 0.999 \\ 
M17 & 2020-06-23 & 38 & 0.14 & 495.37 & 11.71 & 4.03 & 1.57 & -- & 0.94 $\pm$ 0.002 & -- & 0.996 \\ 
M17 & 2020-07-30 & 24 & 0.07 & 117.71 & 10.55 & -0.26 & 2.28 & 0.99 $\pm$ 0.004 & 1.01 $\pm$ 0.002 & 0.995 & 1.0 \\ 
M17 & 2020-09-01 & 21 & 0.05 & 46.45 & 7.17 & -2.16 & -1.24 & 1.03 $\pm$ 0.003 & 1.04 $\pm$ 0.002 & 1.014 & 1.006 \\ 
M17 & 2020-10-10 & 19 & 0.18 & 1104.4 & 13.81 & -2.19 & -1.45 & -- & 0.87 $\pm$ 0.004 & -- & 0.998 \\ 
M17 & 2021-03-04 & 62 & 0.06 & 109.51 & 10.89 & 0.62 & 0.23 & 0.95 $\pm$ 0.004 & 1.0 $\pm$ 0.002 & 0.996 & 0.997 \\ 
M17 & 2021-04-06 & 50 & 0.07 & 89.01 & 8.93 & 0.46 & -0.36 & 1.01 $\pm$ 0.003 & 1.01 $\pm$ 0.002 & 0.998 & 1.005 \\ 
M17 & 2021-05-17 & 68 & 0.06 & 59.7 & 7.7 & 1.12 & -0.64 & 0.94 $\pm$ 0.003 & 1.0 $\pm$ 0.002 & 1.002 & 1.005 \\ 
M17 & 2021-06-14 & 26 & 0.09 & 114.92 & 9.12 & -0.9 & -0.86 & 1.07 $\pm$ 0.005 & 1.02 $\pm$ 0.002 & 0.964 & 0.994 \\ 
M17 & 2021-07-22 & 12 & 0.13 & 355.04 & 11.15 & 0.26 & 1.11 & -- & 0.89 $\pm$ 0.003 & -- & 0.997 \\ 
M17 & 2021-08-22 & 26 & 0.12 & 285.96 & 10.67 & -1.06 & -2.63 & 0.97 $\pm$ 0.016 & 0.98 $\pm$ 0.003 & 1.027 & 1.001 \\ 
M17 & 2021-09-27 & 21 & 0.1 & 152.84 & 9.52 & 0.32 & 5.57 & 0.88 $\pm$ 0.006 & 0.94 $\pm$ 0.002 & 0.978 & 0.999 \\ 
M17 & 2021-11-01 & 13 & 0.15 & 467.54 & 11.76 & 0.66 & 4.75 & -- & 0.9 $\pm$ 0.003 & -- & 1.006 \\ 
M17SWex & 2020-02-22 & 74 & 0.05 & 62.22 & 9.66 & 2.3 & 3.38 & 1.34 $\pm$ 0.007 & 1.18 $\pm$ 0.004 & 0.939 & 0.864 \\ 
M17SWex & 2020-05-21 & 35 & 0.15 & 631.89 & 11.7 & 2.33 & 5.6 & -- & 0.98 $\pm$ 0.004 & -- & 0.997 \\ 
M17SWex & 2020-06-23 & 34 & 0.13 & 408.85 & 10.9 & 7.3 & 8.37 & -- & 0.94 $\pm$ 0.004 & -- & 0.993 \\ 
M17SWex & 2020-07-30 & 29 & 0.07 & 98.37 & 9.52 & 6.48 & 8.2 & 1.01 $\pm$ 0.009 & 1.01 $\pm$ 0.004 & 1.003 & 1.001 \\ 
M17SWex & 2020-09-02 & 16 & 0.07 & 65.66 & 7.55 & 3.76 & 5.05 & 0.84 $\pm$ 0.006 & 1.0 $\pm$ 0.003 & 1.018 & 0.993 \\ 
M17SWex & 2020-10-10 & 15 & 0.15 & 479.55 & 11.73 & 4.62 & 1.68 & -- & 0.84 $\pm$ 0.005 & -- & 0.987 \\ 
M17SWex & 2021-03-03 & 84 & 0.05 & 64.46 & 9.65 & 7.68 & 9.1 & 1.0 $\pm$ 0.005 & 1.04 $\pm$ 0.004 & 1.007 & 1.003 \\ 
M17SWex & 2021-04-06 & 46 & 0.08 & 95.59 & 9.11 & 7.26 & 8.78 & 1.0 $\pm$ 0.01 & 1.04 $\pm$ 0.003 & 1.027 & 1.005 \\ 
M17SWex & 2021-05-17 & 75 & 0.06 & 59.07 & 7.6 & 0.76 & 6.11 & 0.92 $\pm$ 0.007 & 1.0 $\pm$ 0.004 & 0.999 & 1.009 \\ 
M17SWex & 2021-06-14 & 30 & 0.08 & 92.24 & 8.24 & 4.6 & 7.41 & 1.05 $\pm$ 0.009 & 1.05 $\pm$ 0.006 & 1.016 & 0.996 \\ 
M17SWex & 2021-08-04 & 15 & 0.19 & 1199.69 & 14.48 & 8.64 & 7.8 & -- & 0.89 $\pm$ 0.015 & -- & 0.991 \\ 
M17SWex & 2021-09-08 & 24 & 0.15 & 580.55 & 13.1 & 8.02 & 4.3 & -- & 0.96 $\pm$ 0.011 & -- & 0.985 \\ 
M17SWex & 2021-10-08 & 21 & 0.08 & 125.23 & 11.28 & 5.33 & 8.36 & 0.82 $\pm$ 0.033 & 0.95 $\pm$ 0.009 & 0.964 & 0.987 \\ 
S255 & 2020-02-22 & 12 & 0.04 & 56.6 & 9.62 & 1.64 & 0.0 & 0.65 $\pm$ 0.024 & 0.93 $\pm$ 0.005 & 1.107 & 1.022 \\ 
S255 & 2020-02-25 & 14 & 0.05 & 76.64 & 10.15 & 4.98 & -0.03 & 0.98 $\pm$ 0.01 & 1.04 $\pm$ 0.005 & 1.011 & 1.002 \\ 
S255 & 2020-08-11 & 41 & 0.14 & 450.76 & 11.23 & 6.5 & 2.82 & -- & 1.01 $\pm$ 0.006 & -- & 1.003 \\ 
S255 & 2020-09-14 & 49 & 0.07 & 106.15 & 8.15 & 2.53 & 2.78 & 0.96 $\pm$ 0.01 & 0.97 $\pm$ 0.006 & 0.999 & 1.001 \\ 
S255 & 2020-10-31 & 24 & 0.13 & 417.19 & 10.89 & 2.47 & -3.01 & -- & 1.02 $\pm$ 0.006 & -- & 1.003 \\ 
S255 & 2020-12-12 & 58 & 0.15 & 623.63 & 11.7 & 1.75 & -4.15 & -- & 0.94 $\pm$ 0.006 & -- & 0.997 \\ 
S255 & 2021-01-15 & 45 & 0.13 & 357.96 & 10.46 & 3.92 & 0.04 & -- & 1.01 $\pm$ 0.005 & -- & 1.0 \\ 
S255 & 2021-02-12 & 13 & 0.1 & 248.25 & 9.37 & 4.65 & 0.86 & 1.18 $\pm$ 0.018 & 0.99 $\pm$ 0.005 & 0.949 & 1.003 \\ 
S255 & 2021-03-26 & 18 & 0.07 & 144.38 & 9.36 & 3.28 & -1.45 & 1.02 $\pm$ 0.011 & 0.97 $\pm$ 0.006 & 0.99 & 1.005 \\ 
S255 & 2021-08-26 & 36 & 0.06 & 89.42 & 8.86 & 1.87 & 0.41 & 1.15 $\pm$ 0.019 & 1.05 $\pm$ 0.006 & 0.995 & 1.0 \\ 
S255 & 2021-10-08 & 59 & 0.1 & 230.74 & 10.69 & 1.81 & -3.43 & 0.97 $\pm$ 0.028 & 0.97 $\pm$ 0.005 & 0.957 & 0.999 \\ 
S255 & 2021-11-09 & 46 & 0.06 & 73.92 & 7.8 & 2.51 & 1.51 & 0.96 $\pm$ 0.014 & 1.0 $\pm$ 0.007 & 1.007 & 0.99 \\ 
S255 & 2022-01-11 & 35 & 0.15 & 568.44 & 12.05 & 3.74 & -2.99 & -- & 0.96 $\pm$ 0.006 & -- & 1.001 \\ 
S255 & 2022-02-13 & 37 & 0.07 & 131.88 & 9.49 & 1.09 & 1.17 & 1.2 $\pm$ 0.023 & 1.02 $\pm$ 0.008 & 1.014 & 1.01 \\ 
\end{longtable}}
\end{center}



\end{document}